\documentclass[journal,10pt]{IEEEtran}
\usepackage{graphicx}
\usepackage{amsmath,amssymb,amsfonts}
\usepackage{mathtools}
\usepackage{bbm}
\usepackage{cite}
\usepackage{bm}
\usepackage{siunitx}
\usepackage{algorithm}
\usepackage{algorithmic}
\usepackage{stfloats}
\usepackage{subfigure}
\usepackage{color}
\usepackage{xcolor}
\usepackage{amsthm}

\usepackage{multirow}
\usepackage{array}
\usepackage{url}

\usepackage{tabularx,booktabs}

\usepackage{diagbox}

\begin{document}
	\bibliographystyle{IEEEtran}
	\title{
 A Novel Indicator for Quantifying and Minimizing Information Utility Loss of Robot Teams 
 }

	\author{Xiyu~Zhao,
 Qimei~Cui,~\IEEEmembership{Senior~Member,~IEEE}, Wei~Ni, Quan~Z. Sheng,
 Abbas~Jamalipour,~\IEEEmembership{Fellow,~IEEE},
  Guoshun~Nan, 
  Xiaofeng Tao,
  and Ping~Zhang,~\IEEEmembership{Fellow,~IEEE}

  \thanks{Received 30 September 2024; revised 16 January 2025; accepted 28 February 2025. The work was supported by the Joint funds for Regional Innovation and Development of the National Natural Science Foundation of China (Grant No. U21A20449) and the National Key Research and Development Program of China (Grant No. 2020YFB1806804). \textit{(Corresponding authors: Qimei Cui.)}}

  \thanks{X. Zhao, Q. Cui, G. Shun, X. Tao, and P. Zhang are with the Department of
Information and Communication Engineering, Beijing University of Posts and Telecommunications, Beijing 100876, China. 
X. Zhao is also with the School of Computing, Macquarie University, Sydney, NSW 2109, Australia.
Q. Cui, X. Tao, and P. Zhang are also with the Department of Broadband Communication, Peng Cheng Laboratory, Shenzhen 518055, China (e-mail: \{zxy, cuiqimei, nanguo2021, taoxf, pzhang\}@bupt.edu.cn).} 
\thanks{W. Ni and Q. Z. Sheng are with the School of Computing, Macquarie University, Sydney, NSW 2109, Australia. 
(e-mail: 
\{wei.ni, michael.sheng\}@mq.edu.au).}
        \thanks{A. Jamalipour is with the University of Sydney, NSW, Australia. (e-mail: a.jamalipour@ieee.org)}
 
 }
	\maketitle
	\begin{abstract}


The timely exchange of information among robots within a team is vital, but it can be constrained by limited wireless capacity. The inability to deliver information promptly can result in estimation errors that impact collaborative efforts among robots. In this paper, we propose a new metric termed Loss of Information Utility (LoIU) to quantify the freshness and utility of information critical for cooperation. The metric enables robots to prioritize information transmissions within bandwidth constraints.
We also propose the estimation of LoIU using belief distributions and accordingly optimize both transmission schedule and resource allocation strategy for device-to-device transmissions to minimize the time-average LoIU within a robot team. 
A semi-decentralized Multi-Agent Deep Deterministic Policy Gradient framework is developed, where each robot functions as an actor responsible for scheduling transmissions among its collaborators while a central critic periodically evaluates and refines the actors in response to mobility and interference.
Simulations validate the effectiveness of our approach, demonstrating an enhancement of information freshness and utility by 98\%, compared to alternative methods.
	\end{abstract}

	\section{Introduction}

Recent advances in data analytics, artificial intelligence, and robotics have led to the widespread use of autonomous mobile robots, equipped with autonomous navigation and {decision}-making capabilities derived from environmental perception and sensing \cite{luo2022deep}. These robots can apply to manufacturing, logistics, and surveillance, offering advanced functionalities \cite{mu2021intelligent, wu2021multi, hu2022trajectory}. The emergence of robot teams enhances task efficiency and fault tolerance through collaborations~\cite{caillot2022survey, cui2020online}. 

{Consider a warehouse environment where multiple autonomous robots are responsible for picking and transporting items, where device-to-device (D2D) communication is suited for direct communication between nearby robots without relying on a central server or infrastructure. In such a scenario, the robots share their velocity, location, and task intentions in real-time to ensure safe navigation and efficient task allocation~\cite{fu2021survey}. 
Inaccurate or outdated information could lead to inefficient task distribution, potential collisions, or delays in task completion.}
{
{Another application is for search and rescue robot teams, where each robot explores a segment of the disaster area, continuously gathering data from sensors and fellow robots. Each robot integrates this information into its map or knowledge base to prioritize tasks and resource allocations.}}

{{
Wireless networking, especially through D2D communications, further enhances the adaptability and mobility of robots within a team. Robots leverage onboard observations\textcolor{black}{, measurements from other robots,} and D2D links for {decision-making} computation \cite{chen2020wireless, cao2021sum, huang2022edge}.} To ensure the mission accomplishment of robot teams, timely and accurate {operational status} updates are crucial. In scenarios like warehouses with multiple autonomous transportation robots, the prompt sharing of velocity, acceleration, location, and driving intentions is vital for task safety and reliable execution \cite{fu2021survey}.

Limited communication bandwidth among robots may hinder timely {status} transmission. 
{When a robot's exact {operational status} is unavailable to its collaborators, estimation becomes necessary at the collaborators for their informed {decision}-making, e.g., task allocation and path planning, but can distort the status.}
Several metrics have been proposed to evaluate the freshness of information in wireless communication scenarios, with the Age of Information (AoI) being widely adopted \cite{yates2021age}. \textcolor{black}{Although some of the metrics considered the freshness and distortion of information~\cite{mazdin2021distributed}, none was tailored to support the collaboration of teaming robots with a balanced consideration of freshness and distortion.
Thus, there is an urgent need} for a new metric tailored to robot teams, which can effectively prioritize information for transmission {(e.g., the information likely to be distorted by estimation)} in situations where wireless channels face congestion.

{While 5G Ultra-Reliable Low-Latency Communication (URLLC)~\cite{ji2018ultra} and Wireless Time-Sensitive Networking (TSN)~\cite{finn2018introduction} can ensure low latency and high reliability for predefined communication needs, they were not designed for multi-robot collaboration and would typically require dedicated spectrum and centralized coordination.}
{
Multi-agent techniques, such as multi-agent deep reinforcement learning (MADRL),
have been increasingly studied to address scheduling and resource allocation in D2D communication networks~\cite{9770401,10268067,10763434}. However, they generally require centralized training or coordination, and would require all robots to transmit their states (or observations) to the server instantaneously for multi-robot cooperation. The stringent delay requirement of the transmissions could congest the air interface.}
A judicious selection of collaborating robots for transmission becomes crucial, especially in bandwidth-constrained environments.}


This paper presents a new metric called Loss of Information Utility (LoIU) to evaluate the freshness and utility of a robot's {operational status} information needed for its collaborating robots to make effective control {decision}s in a robot team. LoIU can serve as an effective metric for prioritizing the information that a robot needs from its collaborators and, hence, prioritizing the D2D transmissions of the collaborators. {By minimizing the time-average LoIU of a robot team}, we optimize the schedule and resource allocation for the D2D transmissions of critical information {in order to allow the robot} to cooperate effectively. 

The key contributions of this study are listed below:
\begin{itemize}
    \item 
The new metric LoIU is defined to measure the freshness and utility of a robot's {operational status} information in the context of robot teams. LoIU captures the distinct requirements of robots in support of cooperation, quantifying the impact of information freshness and precision.
    \item 
We formulate a new problem to minimize the time-average LoIU of a robot team by optimizing the schedule and resource allocation of the D2D transmissions among the robots. 
The belief distribution is proposed for a robot to estimate LoIU relating to its collaborators. 
    \item 
We interpret the new problem as a decentralized partially observable Markov {decision problem} (Dec-POMDP) with a reward function promoting global performance. 
We design a new MADRL framework to address the Dec-POMDP efficiently in a semi-decentralized fashion.   
 \item 
A new multi-agent deep deterministic policy gradient (MADDPG) algorithm is devised, where each robot acts as an actor to schedule the transmission of its collaborators, and a central critic periodically assesses and refines the actors in response to mobility and interference.

\end{itemize}
{ 
Extensive simulations show that the new LoIU helps improve task reliability by about 33\%, 44\%, 40\%, and 41\%, compared to the latest metrics, i.e., AoI, Age of Incorrect Information (AoII), Urgency of Information (UoI), and Age of Changed Information (AoCI), respectively. 
In coupling with LoIU, the proposed MADRL-based scheduling and resource allocation strategy can effectively improve the freshness and utility of information (i.e., LoIU) by up to 98\%, compared to alternative strategies, including \textcolor{black}{Deep Q-Network (DQN) and Deep Deterministic Policy Gradient (DDPG)}.}  

The subsequent sections are organized as follows: Section II reviews the related work. Section III defines the new metric for a robot team and provides the system model and the formulated problem. In Section IV, we leverage the MADRL method to transform the problem into a Dec-POMDP, and produce an efficient transmission schedule and resource allocation \textcolor{black}{policy} for a robot team. Simulation results validating the efficacy of the designed metric and strategy are presented in Section~V. Section VI provides concluding remarks.
 
\begin{table}[t]
\caption{Notation and definitions}
\label{parameters1}
\centering
\begin{tabular}{|*{2}{l|}}
\hline
\textbf{Notation} & \textbf{Definition} \\ 
\hline
$M$,$J$ & The numbers of mobile robots and resource blocks (RBs) \\
$\cal M$, $\cal J$ & The sets of mobile robots and RBs\\
${\cal C}_m$ & The set of collaborators for robot $m$ \\
${\mathcal{F}}_m(t)$ & The LoIU of robot $m$ at slot $t$ \\
$f_{m,t}^{\mathrm{s}}$, $f_{m,t}^{\mathrm{c}}$ & The {losses} of time utility and content utility of \\ 
 & robot $m$ at slot $t$  \\
$x_{n}^{t}$, $\hat{x}_{n,m}^{t}$ & The actual {operational status} of robot $n$ and its estimate \\ 
 & at robot $m$ in time slot $t$ \\
$e_{n,m}^{t}$ & The estimation error regarding $x_{n}^{t}$ \\
$E_{n,m}$ & The maximum allowed value of $e_{n,m}^{t}$  \\
${\bf X}_{m}^{t}$, ${\bf \hat{X}}_{m}^{t}$ & The sets of $x_{n}^{t}$ and $\hat{x}_{n,m}^{t}$, $\forall n \in {\cal C}_m$ \\
${\bf e}_{m}^{t}$, ${\bf E}_m^t$ & The sets of $e_{n,m}^{t}$ and $E_{n,m}$, $\forall n \in {\cal C}_m$ \\
$\xi_{n,m}^{t}$ & Binary variable that denotes whether $x_{n}^{t}$ is \\
&received by robot $m$ in slot $t$ \\
$q_{n,m}^t$ & Binary variable that decides whether $x_{n}^{t}$ is \\ 
&transmitted to robot $m$ \\
$l_{n,m}^{j,t}$ & Binary variable that decides whether RB $j$ is \\ 
&allocated to the D2D link $(n,m)$ \\
 $\rho_{n,m}^{t}$ &  Binary variable that denotes whether \\ 
 & the transmission of the D2D link $(n,m)$ is successful \\
 $d_{n,m}^t$ & {The transmission delay between $(n,m)$ in slot $t$} \\
 $d_{m}^{t}$ & The delay of {status} updating for robot $m$ in slot $t$ \\
$\nu_{n,m}$ & Binary variable that denotes whether $n \in {\cal C}_m$ \\
$\varsigma_{n,m}^{j, t}$ & The SINR for D2D pair $(n,m)$ {with RB $j$} \\
$\varsigma_{th}$ & The minimum SINR threshold \\
$\alpha_{n,m}^{t}$ & The data size of D2D pair $(n,m)$ at slot $t$ \\
${\bm \alpha}_m^t$ & The set of $\alpha_{n,m}^{t}$ for $\forall n \in {\cal C}_m$ \\
$c_{n,m}^{j, t}$ & The data rate between D2D pair $(n,m)$ {with RB $j$} \\
${\cal M}_{j}^{t}$ & The set of transmitters sharing RB $j$ \\
$g_{m}^t$ & The channel gain between the BS and robot $m$ \\
$g_{n,m}^t$ & The channel gain between $(n,m)$ \\
${\bf S}(t)$ & Global communication state \\
$o_{m}(t)$ & The observation of robot $m$ at slot $t$ \\
$a_{m}(t)$ & The action of robot $m$ at slot $t$ \\
\hline
\end{tabular}
\end{table}

\section{Related Work}
        
\subsection{Scheduling in Robot Teams}	

Many transmission scheduling strategies \textcolor{black}{have been developed recently} for smart factories to ensure transmission reliability \cite{wu2020optimal, vilgelm2017control, lyu2018control, cai2017modulation} and reduce latency \cite{zhou2019access, li2019hybrid,chen2023augmented,noh2021delay,cui2022multi}. In \cite{wu2020optimal}, an optimal machine-to-machine (M2M) transmission scheme was proposed for mission-critical monitoring tasks. It aimed to minimize system {status} estimation errors while satisfying cellular cost and system stability requirements. Vilgelm \textit{et al.}~\cite{vilgelm2017control} developed a cross-layer approach for uplink resource allocation to minimize network errors in support of multiple control systems. Ling \textit{et al.} \cite{lyu2018control} addressed severe mutual interference and transmission collisions by designing a proactive and context-aware approach to optimize channel assignment, power management, and relay collaboration mechanisms. They minimized control-communication costs, adapting to changing channel conditions. In \cite{cai2017modulation}, a modulation-free M2M multi-access protocol was proposed for mission-critical applications, alleviating collisions in wireless channels.


To address delay-tolerant machine-type communication (MTC) within smart factories, Zhou \textit{et al.} \cite{zhou2019access} designed an incentive scheme based on contracts to encourage deferment of access requests by delay-tolerant devices. They devised a long-term cross-layer online resource scheduling method using Lyapunov optimization. In \cite{li2019hybrid}, an adaptive resource scheduling scheme provides real-time computing services in smart manufacturing, employing greedy and threshold strategies with delay constraints. Chen \textit{et al.} \cite{chen2023augmented} formulated a collaborative task offloading framework to minimize the weighted sum of latency and energy requirements. They allocated transmission bandwidth and CPU resources in a multi-hop Mobile Edge Computing (MEC)-based Industrial Internet of Things (IIoT), utilizing an enhanced differential evolution algorithm for efficient problem-solving. Noh \textit{et al.} \cite{noh2021delay} minimized the average task latency in an IIoT network through a series of single-server problems, each solved using convex optimization and MADRL. Addressing delicate deadlines of interdependent computations in an MEC-based robot team, Cui \textit{et al.} \cite{cui2022multi} developed MADRL-based user association and resource allocation to maximize the long-term task completion rate.

{
These previous studies focused on specific objectives of robot teams without considering information exchange between robots (e.g., latency minimization or energy efficiency). There is a scarcity of research on transmission and resource scheduling that are specifically designed to support timely and accurate information updates within robot teams.} Furthermore, the aforementioned studies may not directly apply to a robot team with diverse requirements.

\begin{table*}[]
\caption{  {Comparison of the proposed metric, LoIU, with \textcolor{black}{existing} metrics, where the notation used is defined in Table \ref{parameters1}.}}
\label{parameters0}
    \centering
    \begin{tabular}{c|c|c|c|c|c|c|c}
    \hline
       \textbf{Metrics}  & \textbf{AoI}\cite{kaul2012real} & \textbf{AoS}\cite{zhong2018two} & \textbf{AoII}\cite{maatouk2020age} & \textbf{UoI}\cite{zheng2020urgency} & \textbf{AoCI}\cite{wang2021age} & 
       {\textbf{AoCI}\cite{wang2021minimizing}}&
       \textbf{LoIU} (proposed) \\
       \hline
       \textbf{Definition} &{\tiny$t-U(t)$}&{\tiny$\max\{t-U_{\mathrm{s}}(t),0\}$}&{\tiny$g_{\mathrm{f}}(t)g_{\mathrm{a}}(x^t,\hat{x}^t)$}& {\tiny$\omega(t)g_{\mathrm{a}}(x^t,\hat{x}^t)$}&{\tiny$t\!-\!\min\{S_{i''}|$}&{\tiny $h_{ij}(t)\vartheta^{|\Delta_{ij}(t)|}$}&{\tiny$f_{m,t}^{\mathrm{s}}f_{m,t}^{\mathrm{c}}$}\\
        &{\tiny$U(t)\!=\!\max\{S_{i}|$}&{\tiny $U_{\mathrm{s}}(t)\!=\!\min \{S_{i'}|$}&
        &
        &{\tiny $D_{i'}\!<\!D_{i''}\!\leq\! D_{i}\}$}&&{\tiny$f_{m,t}^{\mathrm{s}}\!=\!{F_{\mathrm{s}}}(d_m^t,{\cal D}_m^t)$}\\
        &{\tiny$D_{i}\!<\!t\}$}&{\tiny$S_{i'}\!>\!D_{i}\}$}&&&{\tiny$D_{i'}=\max\{D_{i'}|$}&&{\tiny$f_{m,t}^{\mathrm{c}}\!=\!{F_{\mathrm{c}}}({\bf X}_m^t,$}\\
        &&&&&{\tiny $x^{S_{i'}}\neq x^{S_{i}}\}$}&&{\tiny ${\bf{\hat X}}_m^t,{\bf E}_m)$} \\
       \hline
        \textbf{Information} & \checkmark & \checkmark & \checkmark & $\times$  & \checkmark & \checkmark  & \checkmark \\
        \textbf{freshness?}& (w/o deadline)& (w/o deadline)&(w/o deadline) & (w/o deadline) &(w/o deadline)&(w/o deadline)& (with deadline) \\
        \hline
        \textbf{Information}  & $\times$ & \checkmark & \checkmark  & \checkmark & \checkmark  & $\times$ & \checkmark  \\
        \textbf{distortion?}&& (w/o threshold)&(w/o threshold)&(w/o threshold) &(w/o threshold)&&(with threshold)\\
        \hline
        \textbf{Collaboration}  & $\times$ & $\times$ & $\times$ & $\times$ & $\times$ &$\times$& \checkmark \\
        \textbf{supported?} &&&&&&&\\
        \hline
    \end{tabular}
\end{table*}

\subsection{Performance Metrics for {Status} Updates}

Several metrics have been proposed to quantify information freshness \cite{kaul2012real, zhong2018two, maatouk2020age, zheng2020urgency,wang2021age,wang2021minimizing}. AoI was initially introduced to measure the time elapsed since the latest received packet was generated~\cite{kaul2012real} and has been applied in different systems \cite{sinha2019scheduling, li2020learning, liu2019age, fang2022age,farazi2018age,vikhrova2020age}. In \cite{sinha2019scheduling}, an energy-efficient greedy sensor scheduling algorithm was proposed to minimize overall AoI, considering the performance criteria of cyber and physical elements in an industrial wireless sensor-actuator network. The authors of \cite{li2020learning} developed a belief-based Bayesian reinforcement learning (RL) approach to optimize energy efficiency with AoI constraints in D2D-enabled industrial wireless networks. In \cite{liu2019age}, two heuristics were devised to address multi-channel allocation in wireless networks to reduce the maximum AoI across all access points. In \cite{fang2022age}, a closed-form expression for users' average peak AoI in three access schemes was derived, and a time slot allocation method was proposed to control the trade-off between the average peak AoI and power consumption. \textcolor{black}{The authors of \cite{farazi2018age} considered a two-hop system with different priorities for {operational status} updates, and obtained closed-form expressions for the peak AoI of non-priority packets and tight bounds of priority packets using Laplace-Stiltjes Transform. In \cite{vikhrova2020age}, bounds were derived for the peak and average age in a multi-source, multi-hop network. Near-optimal scheduling was designed using depth-first search. }

AoI solely captures the temporal deviation of {status} updates without considering {operational status} changes~\cite{sun2019sampling, ornee2021sampling}. To address this limitation, Age of Synchronization (AoS) was proposed, measuring the latency since the time when local {status} information becomes outdated at the receiver \cite{zhong2018two}. In a content caching system with multiple remote sources helping update contents on a server, the authors proposed a frequency allocation mechanism for near-optimal AoS.

Maatouk \textit{et al.} \cite{maatouk2020age} defined AoII, combining an increasing time \textcolor{black}{penalty} with an information \textcolor{black}{penalty} reflecting the disparity between estimated and actual process {status}. They designed an optimal transmission policy to minimize the average AoII for a transmitter-receiver pair operating in an imperfect channel. 
The authors of \cite{maatouk2022age} designed a more general version of AoII with any non-decreasing dissatisfaction function for semantic communications. The authors of \cite{zheng2020urgency} introduced UoI to account for a context-aware weight and a cost associated with {operational status} estimation inaccuracy. A centralized scheduling scheme was developed based on Lyapunov optimization to minimize the average UoI. In \cite{wang2021age}, AoCI was proposed to capture both the passage of time and the change of information content, introducing a binary indicator for whether the received content varies from the previously received update. \textcolor{black}{The authors of \cite{wang2021minimizing} proposed Age of Critical Information (AoCI) by combining the AoI with the critical level difference between the local critical information and the critical information at the server. To minimize the average relative AoCI of mobile clients, an imitation learning-based algorithm was designed.}

As summarized in Table~\ref{parameters0}, {these existing metrics measure freshness or correctness in isolation, {where $g_{\mathrm{f}}(t)$ is an increasing time penalty function, $g_{\mathrm{a}}(x^t,\hat{x}^t)$ is an information penalty function, $h_{ij}(t)$ is the AoI of the $j$-th information category for user $i$ at time slot $t$, $\vartheta$ is a constant value that is greater than one; $\Delta_{ij}(t)$ is the level difference between the local critical level of the $j$-th information category and that on the server; $S_{i}$ and $D_{i}$ are the times when the $i$-th information update is generated and delivered, respectively. These metrics}} do not reflect the distinct requirements of robots in terms of information latency and accuracy. Furthermore, they do not account for the impact of collaborative information on individual robots in a team.


	\section{System Model and Problem statement}


Consider a robot team comprising $M$ robots \textcolor{black}{operating on a time-slotted basis}. Each robot is equipped with computing and {decision}-making capabilities. ${\cal M} = \{1,\cdots,M\} $ collects the indices of the robots. 
${\cal C}_m \subseteq {\cal M}$ is the set of the collaborating robots of robot $m$; that is, each robot $n \in {\cal C}_m$ needs to send its {operational status} (i.e., local measurement), denoted by $x_n^t$, to robot $m$ via a D2D link at times $t=1,\cdots,T$, where $t$ is the index.
 Robot $m$ decides its move based on {$x_n^t$}, $n \in {\cal C}_m$ and its {status} {$x_m^t$}.
 {As in \cite{sun2019sampling,chen2020stochastic}, we assume that $\{x_m^t, \forall m \in {\cal M}\}$ is a collection of independent Wiener processes. }

When a robot has not received {operational status} updates from some of its collaborators \textcolor{black}{in time}, it has to estimate those {status} updates. The estimate, denoted by \textcolor{black}{$\hat{x}_{n,m}^t$, $\forall m \in \cal{M}$, if $x_{n}^{t}$ does not get through to robot $m$ at slot $t$}, can be inaccurate. We develop a new metric, i.e., LoIU, to quantify the relative timeliness and accuracy concerning the estimates, as delineated below.

	\subsection{Loss of Information Utility (LoIU)}
	
 
For any robot $m \in {\cal M}$, {${\bf X}_{m}^{t}$ and ${\bf \hat{X}}_{m}^{t}$ define the actual and estimated {operational status}es of collaborators needed for robot $m$ to execute its task at slot $t=1,\cdots,T$. In other words, ${\bf X}_{m}^{t}$ is the payload sent by other robots to robot $m$, or estimated if not delivered in time}\footnote{In the context of multi-robot cooperation, a robot's \textit{status} refers to its operational status, {e.g., velocity, location, and acceleration,} and a robot's \textit{decision} refers to its next move decided based on its own and its collaborators' statuses. In the context of the studied scheduling design for multi-robot cooperation, a robot's \textit{state} refers to communication state, comprising the estimation errors, date sizes, and channel gains, and its \textit{action} refer to its selection of collaborators and RBs for transmissions determined by the proposed semi-decentralized MADDPG framework.}:
	\begin{subequations} 
	\begin{align}
    & {\bf X}_{m}^{t}=\left\{x_{n}^{t}, \forall n \in {\cal M}, \nu_{n,m}=1\right\} \,; \label{actualstate} \\
    & {\bf \hat{X}}_{m}^{t}=\left\{\hat{x}_{n,m}^{t}, \forall n \in {\cal M}, \nu_{n,m}=1\right\} \,, \label{estimatestate} 
	\end{align}
	\end{subequations}
{where 
$\nu_{n,m}=1$, if robot $n$ is a collaborator of robot $m$; or $\nu_{n,m}=0$, otherwise.}
{It is useful to measure the} utility of ${\bf \hat{X}}_{m}^{t}$ through time utility and content utility.

\textit{Time utility} measures the effect of {time-delay} on information usefulness. The longer the time elapsed, the lower the time utility, {and consequently the larger the time utility loss. We define the time utility loss, denoted by $f_{m,t}^{\mathrm{s}}$, as}
\begin{equation}
\label{time_utility}
    f_{m,t}^{\mathrm{s}}={F_{\mathrm{s}}}(d_m^t,{\cal D}_m^t), 
\end{equation}
{
where $d_m^t$ is the time elapsed since the actual {operational status} ${\bf X}_{m}^{t}$ was yielded, typically obtained through measurements by robots; 
${\cal D}_m^t$ is the maximum allowed delay for {status} update at slot~$t$. 
 $d_m^t$ can be shorter than the duration of a time slot. Here, $d_m^t$ 
is a general concept that can apply broadly across different scenarios, and will be mathematically expressed for the robot team in Section III-B.}
 
\textit{Content utility} measures the 
distortion
between the received or measured (if not received) and actual {operational status}es. The larger the difference, the lower the content utility and, consequently, the larger the content utility loss, denoted by $f_{m,t}^{\mathrm{c}}$, is: 
\begin{equation}
\label{content_utility}
f_{m,t}^{\mathrm{c}}={F_{\mathrm{c}}}({\bf X}_m^t,{\bf{\hat X}}_m^t,{\bf E}_m).
\end{equation}
where ${\bf E}_m$ collects the maximum allowed errors for {status} $x_{n}^{t}$ in ${\bf X}_{m}^{t}$; that is, 
\begin{equation}
    \label{maximumError}
{\bf E}_m=\left\{E_{n,m}, \forall n \in {\cal M}, \nu_{n,m}=1\right\} ,
\end{equation}
where $E_{n,m}$ is the maximum allowed difference between $x_{n}^{t}$ and $\hat{x}_{n,m}^{t}$ in any time slot $t$, and $E_{n,m} \neq 0$. $E_{n,m}$ can differ among the robots, depending on their tasks and {operational status}es. 

As shown in Table~\ref{parameters0}, the LoIU metric of robot $m$ is 
        \begin{equation}
        \label{CoLoIU}
	{\mathcal{F}}_m(t)=f_{m,t}^{\mathrm{s}} f_{m,t}^{\mathrm{c}} .   
       \end{equation}

Next, we define the time utility loss ${F_{\mathrm{s}}}(d_m^t,{\cal D}_m^t)$ and content utility loss $F_{\mathrm{c}}({\bf X}_m^t,{\bf{\hat X}}_m^t,{\bf E}_m^t)$ of robot $m$ at slot~$t$: 
\begin{equation}
	\label{DefLoTU}
	{F_{\mathrm{s}}}({d_m^t,{\cal D}_m^t})= \frac{{d_m^t}}{{{\cal D}_m^t}};
	\end{equation}
%
\begin{equation}
\begin{aligned}
    \label{LossContent}
    {F_{\mathrm{c}}}({\bf X}_m^t,{\bf{\hat X}}_m^t,{\bf E}_m^t) 
    \!=\!\frac{1}{\sum_{n \in {\cal M}} \nu_{n,m}} \sum_{\forall n \in {\cal M},\nu_{n,m}\!=\!1 }\left|\frac{e_{n,m}^{t}}{E_{n,m}}\right|^{2} ,
    \end{aligned}
\end{equation}
where $e_{n,m}^{t}=x_{n}^{t}-\hat{x}_{n,m}^{t}$ is the estimation error of robot $m$ regarding robot $n$'s {operational status} at slot $t$.

We collect the estimation errors of robot $m$ on its collaborators' {statuses} by defining
\begin{equation}
    \label{ErrorSet}
    {\bf e}_{m}^{t}=\left\{e_{n,m}^{t}, \forall m, n \in {\cal M}, \nu_{n,m}=1, m \ne n\right\} .
\end{equation}

	\subsection{LoIU for Robot Team}
 
	Each robot updates its {operational status} and makes a decision within a time slot, e.g., through an appropriate design of time slot duration and the transmission schedules of the (collaborating) robots. In this way, $d_m^t$ is the overall delay for robot $m$ to update the operational status ${\bf X}_m^t$ in slot~$t$.

Let $\xi_{n,m}^{t}=q_{n,m}^{t} \rho_{n,m}^{t}$ indicate whether the measurement of robot $n$ is received by robot $m$ in slot $t$. $q_{n,m}^t=1$ if robot $n$ transmits to robot $m$, and $q_{n,m}^t=0$, otherwise. $\rho_{n,m}^{t}=1$, if the transmission is successful, and $\rho_{n,m}^{t}=0$, otherwise. As a result, $\xi_{n,m}^{t}=1$, if the measurement of robot $n$ is received by robot $m$ in slot $t$; otherwise, $\xi_{n,m}^{t}=0$. 

{ 
The Wiener process is traditionally defined in continuous-time settings, and has independent, normally distributed increments with zero mean and a variance proportional to the time elapsed~\cite{sun2019sampling}. We discretize the Wiener process by requiring all robots to measure their {operational status}es at the beginning of every time slot $t$, $\forall t=0, 1, \ldots, \infty.$
Let $\varepsilon_{n}(\tau)=x_{n}^{t+\tau}-x_{n}^{t}$ denote the change of error over $\tau$ slots from time slot $t$ to time slot $(t+\tau)$:
\begin{equation}
    \label{gaussianVariable}
    \varepsilon_{n}(\tau) \sim \mathcal{N}\left(0, \sigma_{n}^2 \tau\right) ,
\end{equation}
where $\sigma_{n}^{2}$ is the variance of the increments in a time slot; in other words,
the change between two consecutive time slots (i.e., $x_{n}^{t+1}-x_{n}^{t}$, $\forall t$) is an i.i.d. Gaussian random variable with zero mean and variance~$\sigma_{n}^{2}$.}

If robot $m$ does not receive robot $n$'s {operational status} at slot $t$ (i.e., $\xi_{n,m}^{t}=0$), it takes the (estimated) {status} of robot $n$ at slot $(t-1)$ as the estimated {status} at slot $t$; i.e., 
	\begin{equation}
	\label{EstimateState}
	\hat{x}_{n,m}^{t}=\xi_{n,m}^{t} x_{n}^{t}+\left(1-\xi_{n,m}^{t}\right) \hat{x}_{n,m}^{t-1} .
	\end{equation}
According to (\ref{EstimateState}), the estimation error of ${x}_{n}^{t}$ is given by
	\begin{equation}
        \begin{aligned}
	\label{EstimateError}
	e_{n,m}^{t}\!=\!&(1\!\!-\!\xi_{n,m}^{t})(x_{n}^t\!\!-\!\hat{x}_{n,m}^{t-1}) 
        \!=\!\left(1\!-\!\xi_{n,m}^{t}\right)\left(\varepsilon_{n}^{t}\!+\!e_{n,m}^{t-1}\right) , 
        \end{aligned}
	\end{equation}
where $\varepsilon_{n}^{t}\!=\!x_{n}^t-{x}_{n}^{t-1}$ is the change in the error ${e}_{m n}$ at slot $t$.

The delay encompasses queuing delay, processing delay, and propagation delay. The processing delay remains constant. The queuing delay is negligible, considering delay-sensitive tasks. Thus, our focus is on the propagation delay, $d_{m}^{t}$, of updating the {statuses of robot $m$'s collaborators}, 
as given by
\begin{equation}
    \label{UpdateDelay}
    d_{m}^{t}=\max _{n \in {\cal M}}\left\{d_{n,m}^{t} |  \nu_{n,m}=1, q_{n,m}^t=1 \right\} ,
\end{equation}
where $d_{n,m}^t$ is the delay in transmitting the measurement from robot $n$ to robot $m$. If $f_{m,t}^{\mathrm{s}} \le 1$, then $d_{m}^{t} \leq {{\cal D}_m^t}$. 

According to {(\ref{CoLoIU})}, (\ref{DefLoTU}), and (\ref{LossContent}), the LoIU of all {statuses} updated by robot $m$ is designed as
\begin{equation}
\begin{aligned}
    \label{LoIUm}
    \mathcal{F}_{m}(t)=\frac{d_{m}^{t}}{{\cal D}_{m}^{t}} \cdot \frac{1}{\sum_{n \in {\cal M}} \nu_{n,m}} \sum_{\nu_{n,m}=1, \forall n \in {\cal M}}\left|\frac{e_{n,m}^{t}}{E_{n,m}}\right|^{2}  \,.
    \end{aligned}
\end{equation}

	\subsection{Network Model for Robot Team}

 
 Situated in the coverage of  a cellular base station (BS), 
 the $M$ robots reuse the downlink spectrum of the cellular network. Orthogonal frequency division multiple access (OFDMA) is adopted for the D2D communications among the robots, where ${\cal J} = \{1,\cdots,J\} $ collects the indices of the resource blocks (RBs) available in a time slot. Each RB has the bandwidth $W$ and lasts a slot. Multiple D2D links can share the same RB, as long as their interference is tolerable. 
 
 Let $(n,m)$ denote a D2D pair with robot $n$ sending its measurement to robot $m$. 
{We account for background cellular traffic, modeled as a continuous data stream from the BS~\cite{bazzi2021design,rajalakshmi2024towards}, which may introduce interference to the robots. The receiver, i.e., robot $m$, is subject to interference not only from the BS but also from transmitters of other D2D links that share the same RBs as the D2D pair $(n,m)$.}

Let $l_{n,m}^{j,t}$ indicate whether RB $j$ is allocated to D2D pair $(n,m)$ at the $t$-th slot. $l_{n,m}^{j,t}=1$, if the RB is allocated to the D2D pair; $l_{n,m}^{j,t}=0$, otherwise. 
{At slot $t$, the transmitters of the D2D pairs sharing RB $j$ are collected by
\begin{equation}
    \label{SharingRBset}
    {\cal M}_{j}^{t}=\left\{n \mid l_{n,m}^{j, t}=1, \forall m, n \in {\cal M}\right\} \,.
\end{equation}}
Then, the Signal-to-Interference-plus-Noise Ratio (SINR) at robot $m$ in D2D pair $(n,m)$ is
\begin{equation}
    \label{SINRmj}
    \varsigma_{n,m}^{j, t}=\frac{P_{r} g_{n,m}^{t}}{P_{b} g_{m}^{t}+\sum_{n' \in {\cal M}_{j}^{t}, n' \neq n} P_{r} g_{n',m}^{t}+\sigma^{2}} \,,
\end{equation}
where $P_r$ and $P_b$ are the transmit powers of the robots and BS, respectively; $g_{n,m}^t$ is the channel gain between D2D pair $(n,m)$ in the $t$-th slot; $g_{m}^t$ and $g_{n',m}^t$ are the channel gains of the interference links from the BS to robot $m$ and from robot $n'$ to robot $m$ in slot $t$, respectively; $\sigma^{2}$ is the power of {additive white Gaussian noise (AWGN)}. 

The data rate of D2D pair $(n,m)$ is given by
\begin{equation}
    \label{dataRate}
    c_{n,m}^{j, t}=W \log_2 \left(1+\varsigma_{n,m}^{j, t}\right).
\end{equation}
The transmission delay from robot $n$ to $m$ can be written~as
\begin{equation}
    \label{transdelaymj}
    d_{n,m}^{t}=\frac{\alpha_{n,m}^{t}}{c_{n,m}^{j, t}} \,,
\end{equation}
where $\alpha_{n,m}^{t}$ is the data size (in bits) of the D2D pair at slot~$t$.

Robot $m$ can receive the measurement from robot $n$ only if an RB is allocated to D2D pair $(n,m)$ {(i.e., $\sum_{j=1}^{J} l_{n,m}^{j,t}=1$)} and the transmission is successful {(i.e., $\rho_{n,m}^t=1$)}. Let $\varsigma_{th}$ be the minimum required SINR threshold; i.e., $\rho_{n,m}^t=1 $ if $\varsigma_{n,m}^{j,t} \geq \varsigma_{\rm th}$; $\rho_{n,m}^t=0$, otherwise.

\subsection{Problem statement}
We propose to minimize the average LoIU of the robot team, while satisfying the constraints of latency, i.e.,{
\begin{equation}
\label{optimizationP}
    \begin{aligned}
\textbf{P}:  \min \limits_{\{\boldsymbol{a}(t), \forall t\}} &\lim _{T \to \infty} \frac{1}{M T} \sum_{t=0}^{T-1} \sum_{m=1}^{M} \left( f_{m,t}^{\mathrm{s}} \cdot f_{m,t}^{\mathrm{c}} \right) \\
\textrm {s.t. }  \quad&\textbf{C1}: f_{m,t}^{\mathrm{s}} \leq 1, \forall m \in {\cal M} \,, \\
& \textbf{C2}: {\sum_{j=1}^{J} l_{n,m}^{j,t} \leq 1 \,, \forall m, n \in {\cal M}} ,\\
& \textbf{C3}: l_{n,m}^{j, t} \in\{0,1\}, \forall j \in {\cal J}, \forall m, n \in {\cal M} \,, 
    \end{aligned}
\end{equation}
where ${\bf a}(t)\!=\!\{l_{n,m}^{j,t}, \forall {m,n} \!\in \!{\cal M},j \!\in \!\!{\cal J}, n \!\neq \!\!m,  \nu_{n,m}\!=\!1  \}$ collects all optimization variables} concerning collaborator selection to transmit and allocation of the corresponding resources at slot $t$. \textbf{C1} imposes {a constraint on} delivery latency. \textbf{C2} specifies {that} at most one RB is allocated to each D2D pair. \textbf{C3} is self-explanatory. 

{While we do not explicitly enforce a constraint on the estimation error, it has been incorporated into the optimization objective, which is captured by the proposed LoIU.
Serving as the optimization objective, LoIU captures both freshness and content accuracy, ensuring the robots prioritize transmitting the most relevant and fresh information in a resource-constrained environment. By optimizing LoIU, we can enhance the freshness and content utility of shared information, directly improving collaboration efficiency and task success.}

Problem \textbf{P} is difficult to solve directly for three reasons. First, the SINR of a D2D link is determined by the RB selection of the other D2D {pairs}, which, in turn, affects the LoIU of a robot. 
Second, the {optimization variables} of all robots are coupled. Last but not least, for each D2D pair, the selection of transmitting robots and RBs is part of the {variables}. Specifically, the problem is an integer program with non-linear constraints and can be reduced to 0–1 integer programming, which is one of Karp's 21 NP-complete problems~\cite{karp2010reducibility}. As a result, problem \textbf{P} is an NP-hard problem~\cite{bellare1995complexity}. To cope with these challenges, we design semi-decentralized scheduling and resource allocation using MADRL in the following sections.

\section{Proposed Semi-Decentralized MADRL}
In the face of time-varying channel gains and data sizes, we propose a new MADRL-based scheduling and resource allocation algorithm where the robots serve as agents. The average LoIU of the system is minimized by enabling the agents to adjust their policies in a semi-decentralized manner.

	\subsection{Problem Interpretation with Dec-POMDP}

Ideally, the estimation errors of the {status}es are known precisely, and resources are allocated centrally based on the latest {communication states} of the robots available at the BS, including the global {communication state} ${\bf S}(t)=\left\{ {\bf e}_m^t, {\bm \alpha}_m^t, {\bf G}_m^t, \forall m \in {\cal M} \right\}$. ${\bf e}_m^t=\left\{e_{n,m}^t, \forall n \in {\cal M} \right\}$. ${\bm \alpha}_m^t=\left\{ \alpha_{n,m}^t, \forall n \in {\cal M} \right\}$. ${\bf G}_m^t=\left\{g_{n,m}^t,g_m^t, \forall n \in {\cal M} \right\}$. {The {action} is ${\bf a}(t)=\left\{ l_{n,m}^{j,t}, \forall {m,n} \in {\cal M}, j \in {\cal J}, n \neq m, \nu_{n,m}=1  \right\}$.} However, centralized {action}-choosing 
incurs significant communication overhead and delays. 

Problem \textbf{P} can be reformulated as a Dec-POMDP model, represented by the tuple $({\cal M},{\cal S},{\cal O}_m,{\cal A}_m,{\cal P},{\cal R})$, $\forall m \in {\cal M}$, in a multi-agent environment. ${\cal S}$ is the global {state} space. ${\cal O}_m$ and ${\cal A}_m$ are the observation and {action} {spaces} of robot $m$, respectively. ${\cal O}_m$ contains only its local information. ${\cal P}$ and ${\cal R}$ denote the joint transition and reward, respectively. At time slot $t$, each robot $m$ takes an {action} on its observation $o_m(t)$ and the individual policy $\pi_{m}: o_{m}(t) \rightarrow a_{m}(t)$. After the {action}s of all $M$ robots, the communication {state} ${\bf S}(t)$ transits to a new {state} ${\bf S}(t+1)$. 
The Dec-POMDP model defines the {communication state}, observation, {action}, and reward below:

$\bullet$ \textbf{State}: 
{The communication {state} consists of} the estimation errors, data sizes, and channel gains between the robots and from the BS to the robots, as given by 
\begin{equation}
    \label{agentstate}
    {\bf S}(t)\!=\!\left\{e_{n,m}^{t}, \alpha_{n,m}^{t}, g_{n,m}^{t}, g_{m}^{t}, \forall n,m \!\in \!\!{\cal M}, n \!\neq m \!\right\} .
\end{equation}

$\bullet$ \textbf{Observation}:
For robot $m$, the observation is given by
\begin{equation}
    \label{agentObeservation}
    o_{m}(t)\!=\!\left\{ \xi_{n,m}^{t-1}, \alpha_{n,m}^{t}, g_{n,m}^{t}, g_{m}^{t},
    \forall n\! \in \!\!{\cal M}, n \!\neq \!m, \nu_{n,m}\!\!=\!1 \!\right\} ,
\end{equation}
where $\xi_{n,m}^{t-1}$ indicates whether robot $m$ received successfully the {operational status} information from robot $n$ at slot $(t-1)$. 

$\bullet$ \textbf{Action}:
The {action} of robot $m$ at time slot $t$ involves selecting its collaborators to transmit and assigning RBs to the collaborators, i.e., 
\begin{equation}
    \label{agentAction}
    { a_{m}(t)\!\!=\!\left\{l_{n,m}^{j,t}  , \forall n \!\in \!\!{\cal M},j \!\in \!\!{\cal J},  n \!\neq \!m, \nu_{n,m}\!\!=\!\!1 \right\} .}
\end{equation}

$\bullet$ \textbf{Reward}:
Problem \textbf{P} aims to minimize the average LoIU of the robot team, given the available RBs. At slot $t$, the instant local reward of robot $m$ is
\begin{equation}
    \label{localReward}
    r_{m}^{\mathrm{l}}(t)=r_{m}^{\mathrm{L}}(t)+r_{m}^{\mathrm{s}}(t) \,,
\end{equation}
where $r_{m}^{\mathrm{L}}(t)$ accounts for the instant LoIU at slot $t$:
\begin{equation}
    \label{localReward1}
    r_{m}^{\mathrm{L}}(t)=-{\cal F}_{m}(t).
\end{equation}

Based on \textbf{C1}, $r_{m}^{\mathrm{s}}(t)$ is a penalty function and is given by
\begin{equation}
    \label{localReward2}
    { r_{m}^{\mathrm{s}}(t)=-H\left(d_{m}^{t}-{\cal D}_{m}^{t}\right) f_{m,t}^{\mathrm{s}} \epsilon,}
\end{equation}
where $H(\cdot)=1$, if the input is greater than zero; $H(\cdot)=0$, otherwise. $H(\cdot)$ assesses if the latency of the {operational status} updates exceeds the maximum allowed latency (i.e., the loss of time utility $f_{m,t}^{\mathrm{s}}>1$). 
{$\epsilon$ is the penalty coefficient when the maximum latency is exceeded, which takes a large positive value to penalize a robot when its {action} violates~\textbf{C1}. }

To optimize the global system performance and encourage cooperation among the robots, each robot $m \in {\cal M}$ evaluates the global reward, which is defined as
\begin{equation}
    \label{globalreward}
    r_{m}(t)=\frac{1}{M} \sum_{m'=1}^{M} r_{m'}^{\mathrm{l}}(t) \,.
\end{equation}
As a result, problem \textbf{P} is interpreted as a Dec-POMDP, {where the average LoIU of all robots is optimized by maximizing the average sum of their local rewards.} 
Constraint \textbf{C1} is satisfied with the penalty in (\ref{localReward2}), i.e., the global reward $r_m(t)$ in (\ref{globalreward}).

The goal of MADRL is that each robot learns a policy for maximizing its long-term expected return, which is the expected cumulative discounted reward, as given by
\begin{equation}
    \label{MADRLgoal}    J_m=\mathbb{E}\left[\sum_{k=0}^{\infty}{\gamma}^k r_m(k)\right] \,,
\end{equation}
where $\gamma \in \left[0,1\right)$ is the discount factor for the reward, {and $k$ is the time slot increment from time slot $t=0$.}

\subsection{Belief Distribution-Based LoIU Estimation}
According to (\ref{localReward})--(\ref{localReward2}), each robot evaluates its local reward (i.e., LoIU and penalty) after taking an {action} at a slot. 
  {While the exact estimation errors of its collaborators' {status}es are unavailable at a robot until the {operational status}es are received, the expected reward can be obtained and used for reward evaluation \cite{chen2022artificial}, i.e., }
\begin{equation}
    \label{expectedrewards_ref}
    r\left(s,a\right)=\sum_{r\in \cal{R}}r\sum_{s'\in \cal{S}}\Pr\left (s',r|s,a\right ) \,,
\end{equation}
where $\Pr(s',r|s,a)$ is the conditional probability distribution of {communication state} $s'$ and reward $r$, conditioned on {state} $s$ and {action} $a$.

According to {(\ref{localReward}), (\ref{globalreward}), and (\ref{expectedrewards_ref}),} the expected reward for an {action} taken by robot $m$ at slot $t$ is
\begin{equation}
    \label{expected_reward}
    \begin{aligned}
    \overline{r_m\left(t\right)}
    &=\mathbb{E}\left[\frac{1}{M} \sum_{m'=1}^{M} r_{m'}^{\mathrm{l}}(t) \right] 
    =\overline{r^{\mathrm{L}}(t)}+\overline{R^{\mathrm{s}}(t)} ,
        \end{aligned}
\end{equation}
where 
\begin{align} \label{average_reward1}
    &\overline{r^{\mathrm{L}}(t)}\!\!=\!\mathbb{E}\!\left[\frac{1}{M} \!\!\sum_{m'=1}^{M} \!r_{m'}^{\mathrm{L}}(t) \right]; 
    \overline{R^{\mathrm{s}}(t)}\!\!=\!\mathbb{E}\!\left[\frac{1}{M} \!\!\sum_{m'=1}^{M} \!r_{m'}^{\mathrm{s}}(t)\right]. 
    \end{align}

 According to (\ref{localReward2}), $r_{m'}^s(t)$ depends only on $s$ and $a$, and (\ref{average_reward1}) can be assessed based on (\ref{LoIUm}), i.e.,
\begin{subequations}
    \label{expected_reward2}
    \begin{align}
    \overline{R^{\mathrm{s}}(t)}&=\frac{1}{M} \sum_{m'=1}^{M} r_{m'}^{\mathrm{s}}(t) \,;
    \label{expected_reward2_a} \\
    \begin{split}
    \overline{r^{\mathrm{L}}(t)}&=\mathbb{E}\left[-\frac{1}{M} \sum_{m=1}^{M} \sum_{\nu_{n,m}=1, \forall n \in {\cal M}} Z_{n,m}^t \left|e_{n,m}^t \right|^2 \right] \\
    &=-\frac{1}{M} \sum_{m=1}^{M}\mathbb{E}\left[ {\cal F}_{m}(t) \right] =\frac{1}{M} \sum_{m=1}^{M}\overline{r_{m}^{\mathrm{L}}(t)}\,,
    \label{expected_reward2_b}
    \end{split}
    \end{align}
\end{subequations}
where $Z_{n,m}^t$ is a constant for D2D pair $(n,m)$ at slot $t$ after robot $m$ takes an {action}, and is given by
\begin{equation}
\label{Z_mj}
    Z_{n,m}^t=\frac{d_{m}^{t}}{{\cal D}_{m}^{t}} \cdot \frac{1}{\sum_{n \in {\cal M}} \nu_{n,m}} \cdot \frac{1}{\left|E_{n,m}\right|^2} \,.
\end{equation} 
With (\ref{expected_reward}) and (\ref{expected_reward2}), each robot needs its expected LoIU, $\overline{r_{m}^{\mathrm{L}}(t)}$, for evaluation of the overall system reward, $\overline{r_m\left(t\right)}$.

We propose that, at slot $t$, robot $m$ evaluates the expected LoIU {after taking an {action}}, according to its observed $\xi_{n,m}^{t}$ ($\forall n \in {\cal M}$) and the belief distribution of the estimation errors concerning its D2D pair $(n,m)$, $\forall n \in {\cal M}$ and $n \neq m$. A belief represents the imperfect knowledge of a robot regarding the environment when the direct measurement of the {communication state} is impossible. 
The belief distribution assigns each possible {state} with a probability, which is the posterior probability of the {state} variables given the observations \cite{thrun2002probabilistic}. 

The belief distribution of the estimation error $e_{n,m}^t$ is a conditional probability distribution conditioned on the observed $\xi_{n,m}^{t}$ and the estimated {operational status} $\hat{x}_{n,m}^{t}$. The belief distribution of $e_{n,m}^t$, denoted by $b\left(e_{n,m}^{t}\right)$, is given by
\begin{equation}
    \label{beliefDistribution}
    b\left(e_{n,m}^{t}\right)=f\left(e_{n,m}^{t} \mid \xi_{n,m}^{1: t}, \hat{x}_{n,m}^{1:t}\right) ,
\end{equation}
where $f(\cdot|\cdot,\cdot)$ is the conditional probability density function (PDF); $\xi_{n,m}^{1: t}=\{\xi_{n,m}^{1},\cdots,\xi_{n,m}^{t}\}$ is the set that collects whether robot $m$ successfully received the {operational status} information from robot $n$ from slot~1 to slot $t$; $\hat{x}_{n,m}^{1:t}=\{\hat{x}_{n,m}^{1},\cdots,\hat{x}_{n,m}^{t}\}$ is the set that collects the estimated {status} from slot 1 to slot $t$.

The average estimation error $e_{n,m}^t$ can be obtained according to the belief distribution, as given by
\begin{equation}
    \label{averageError}
    \overline {e_{n,m}^t}  = \int_{-\infty}^{\infty} {b(e_{n,m}^t) \cdot e_{n,m}^{t}d{e_{n,m}^t}}.
\end{equation}
We evaluate the expected LoIU based on (\ref{beliefDistribution}) and (\ref{averageError}), i.e.,
\begin{equation}
    \label{estimateLoIU}
    \overline{\mathcal{F}_{m}(t)}=\frac{d_{m}^{t}}{\mathcal{D}_{m}^{t}} \cdot \frac{1}{\sum_{n \in {\cal M}} \nu_{n,m}} \sum_{\nu_{n,m}=1, \forall n \in {\cal M}} \left|\frac{\overline{e_{n,m}^{t}}}{E_{n,m}}\right|^{2} \,.
\end{equation}


The belief distribution of the estimation error $e_{n,m}^t$ has the following cases:
\begin{itemize}
    \item When $\xi_{n,m}^t=1$, the {operational status} $x_{n}^t$ is updated at slot $t$, i.e.,
    \begin{align}
        &e_{n,m}^{t}\!=\!x_{n}^{t}-\hat{x}_{n,m}^{t}\!=\!0 ; \\
       &b\left(e_{n,m}^{t}\!=\!0 \!\mid \!\xi_{n,m}^{t}\!=\!1\right)\!=\!1 . \label{updateBelief}
    \end{align}
    \item When $\xi_{n,m}^t=0$, $x_{n}^t$ is not updated at slot $t$, i.e.,
    \begin{equation}
        \label{notupdateError}
\begin{aligned}
e_{n,m}^{t} &=x_{n}^{t}-\hat{x}_{n,m}^{t}=x_{ n}^{t}-\hat{x}_{n,m}^{t-1} \\
&=\left(x_{n}^{t}-x_{n}^{t-1}\right)+\left(x_{n}^{t-1}-\hat{x}_{n,m}^{t-1}\right)  \\
&=\varepsilon_{n}(1)+e_{n,m}^{t-1}  \,.
\end{aligned}
    \end{equation}
According to (\ref{gaussianVariable}), $e_{n,m}^{t}$ is a Gaussian random variable:
    \begin{equation}
    \label{errorgaussian}
    e_{n,m}(t) \sim \mathcal{N}\left(0, \tau_{n,m}^t\sigma_{n}^{2} \right) \,,
\end{equation}
where $\tau_{n,m}^t$ is the number of slots since the last time $x_{n}^{t_0}, \forall t_0 \leq t$, was received by robot $m$. At slot $t$, $\tau_{n,m}^t=\tau_{n,m}^{t-1}+1$ if $x_{n}^t$ is not received; otherwise, $\tau_{n,m}^t=0$. The belief distribution of $e_{n,m}^t$ in (\ref{notupdateError}) is 
\end{itemize}
\begin{equation}
       b\left(e_{n,m}^{t} \!\mid \!\xi_{n,m}^{t}\!=\!0\right)\!=\!\frac{1}{\sqrt{2 \pi \tau_{n,m}^{t} \sigma_{n}^{2}}} \exp \left[-\frac{\left(e_{n,m}^{t}\right)^{2}}{2 \tau_{n,m}^{t} \sigma_{n}^{2}}\right] \label{emjkBelief}. 
\end{equation}
For conciseness, {$b\left(e_{n,m}^{t} \mid \xi_{n,m}^{t}\right)$ is rewritten as $b_{n,m}^t$. }The expected LoIU of robot $m$ at slot $t$ in (\ref{estimateLoIU}) is rewritten as
\begin{equation}
    \label{estimateLoIUm}
    \begin{aligned}
\overline{\mathcal{F}_{m}(t)} 
&\!=\!\frac{d_{m}^{t}}{{\cal D}_{m}^{t}} \!\cdot \!\frac{1}{\sum_{n \in {\cal M}} \nu_{n,m}} \sum_{n \in {\cal M}} \frac{\nu_{n,m}}{|E_{n,m}|^{2}} \mathbb{E}_{e_{n,m}^{t} \!\sim \!b_{n,m}^{t}}\!\big[|e_{n,m}^{t}|^{2}\big] \\
&=\frac{d_{m}^{t}}{{\cal D}_{m}^{t}} \cdot \frac{1}{\sum_{n \in {\cal M}} \nu_{n,m}} \sum_{n \in {\cal M}} \frac{\nu_{n,m} \left(1-\xi_{n,m}^{t}\right) }{\left|E_{n,m}\right|^{2}} \tau_{n,m}^{t} \sigma_{n}^{2} \,,
\end{aligned}
\end{equation}
{where $\mathbb{E}_{e_{n,m}^{t} \sim b_{n,m}^{t}}\left[\cdot \right]$ takes expectation conditioned on the belief distribution of $e_{n,m}^t$, i.e., $b_{n,m}^t$ in \eqref{updateBelief} or~\eqref{emjkBelief}.}

Finally, by substituting \eqref{estimateLoIUm} into (\ref{localReward1}), $r_{m}^{\mathrm{L}}(t)$ is rewritten as 
\begin{equation}
    \label{estimateReward1}
    r_{m}^{\mathrm{L}}\!(t)\!\!=\!-\frac{d_{m}^{t}}{{\cal D}_{m}^{t}} \frac{1}{\sum_{n \in {\cal M}} \nu_{n,m}} \sum_{n \in M} \!\frac{\nu_{n,m} \!\!\left(1\!\!-\!\xi_{n,m}^{t}\right) }{\left|E_{n,m}\right|^{2}} \tau_{n,m}^{t} \sigma_{n}^{2} .
\end{equation}


\subsection{Semi-Decentralized MADDPG for Robot Scheduling}

 \begin{figure}[t]
\centering  
\subfigure[The architecture of the traditional MADDPG framework]{
\label{fig:tra-maddpg}
\includegraphics[width=0.45\textwidth]{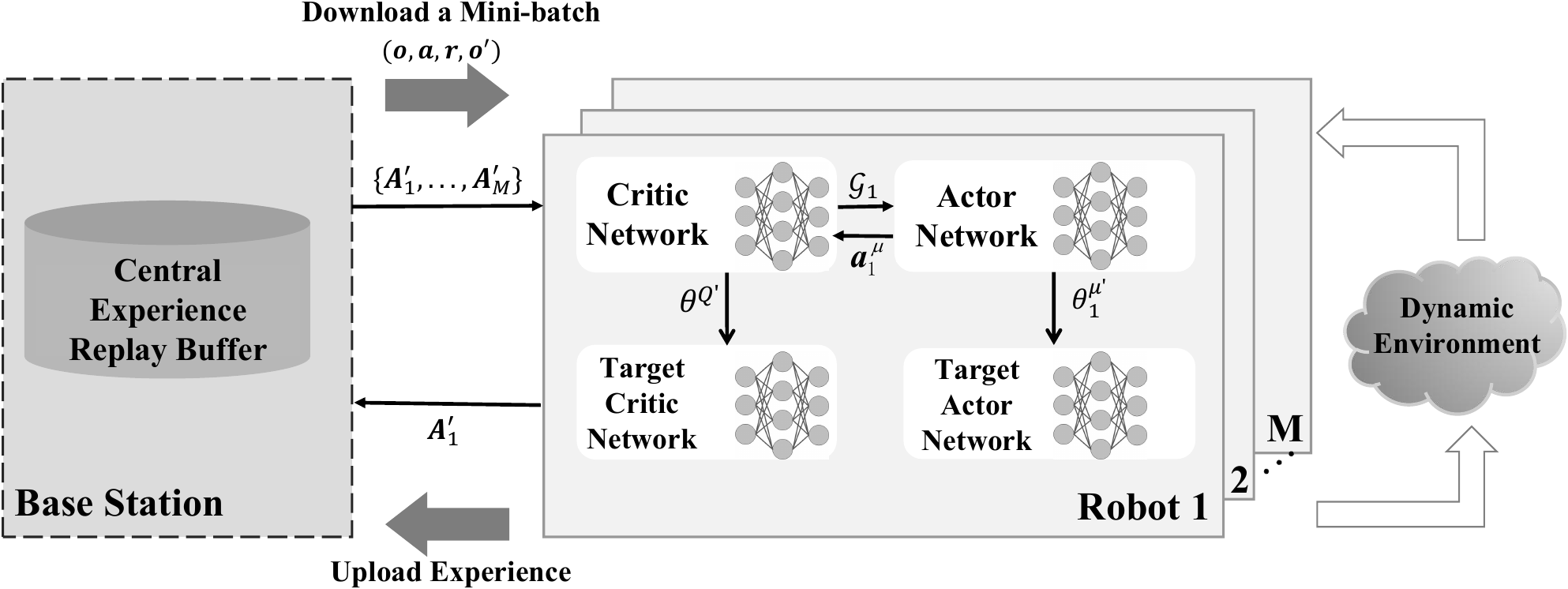}}
\\
\subfigure[The architecture of the proposed semi-decentralized MADDPG framework for semi-decentralized training and decentralized execution. Each robot has an actor network, a target actor network, and a local experience replay buffer. The BS is equipped with a critic network, a target critic network, and a central experience replay buffer.]{
\label{fig:maddpg}
\includegraphics[width=0.45\textwidth]{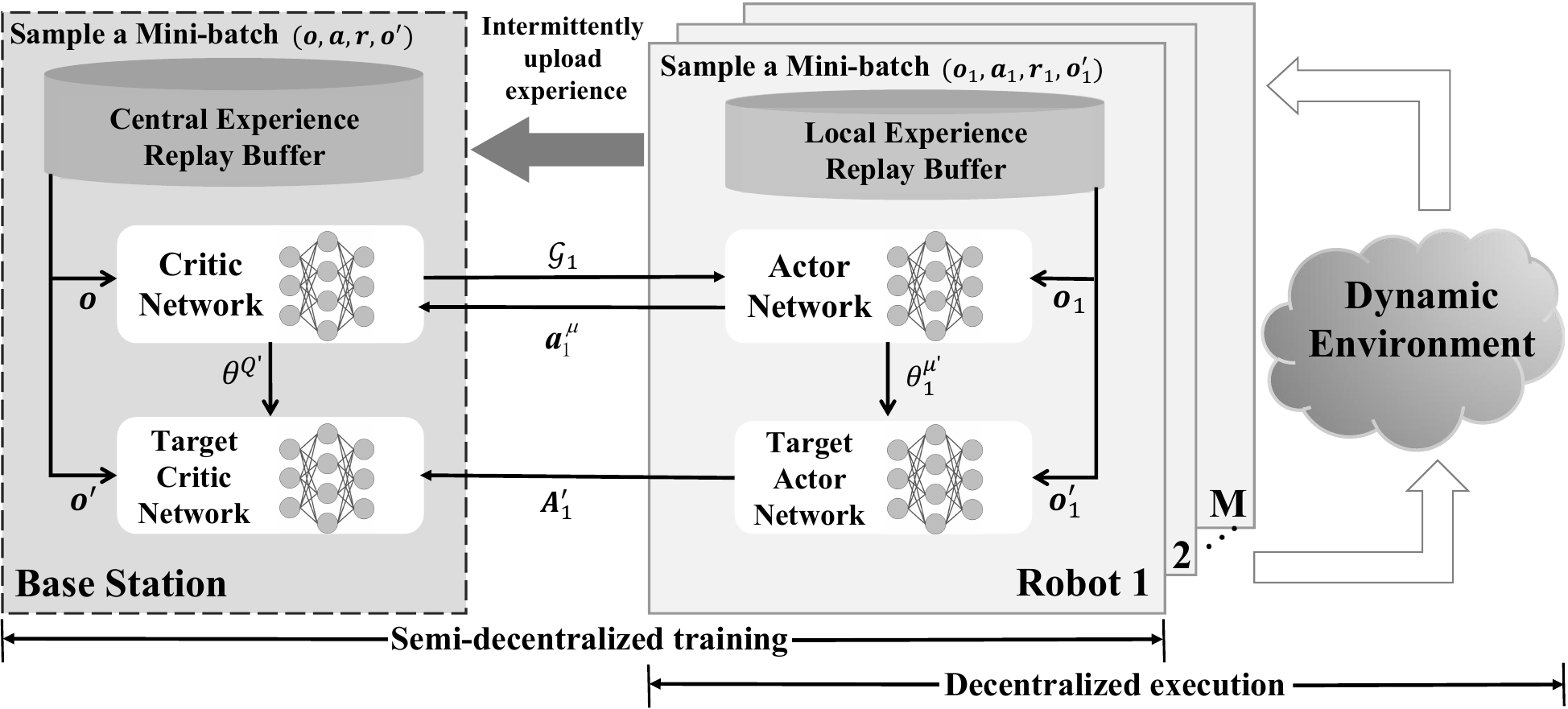}}
\caption{ {The comparison of the proposed semi-decentralized MADDPG vs, the traditional MADDPG framework.}}
\label{loss-t}
\end{figure}
 
{To circumvent stringent communication delay constraints during the training process to support multi-robot collaboration in resource-constrained environments, a} novel semi-decentralized MADDPG framework is designed to resolve the Dec-POMDP problem in (\ref{optimizationP}). {Given the common reward, the} MADDPG model allows the robots to decide collaboratively on collaborator selection and resource allocation in a semi-decentralized fashion.
{Each robot, acting as an agent, only upload its {communication state} (or experience) to enrich the replay memory at the server with much relaxed delay constraints. Moreover, a}
 relatively small number of parameters need to be exchanged between the BS and robots at each training step, reducing communication overhead. { During execution, an agent can independently choose its {action} based on its local observation. By contrast, a centralized implementation of DDPG/MADDPG {would require all robots to instantaneously transmit their {communication state}s (and observations) to the server.} The stringent delay requirement of the transmissions could lead to congestion in resource-constrained air interfaces.}

 {As shown in Fig.~\ref{fig:tra-maddpg}, the} traditional MADDPG algorithm is characterized by centralized training and distributed execution, where every agent possesses {an actor network, a target actor network, a critic network, and a target critic network.} Every agent makes a {decision} through its actor network, based on its local observation. The critic network of the agent evaluates the {action} based on all agents' experience. After carrying out {action}s ${\bf a}(t)$ in {state} ${\bf S}(t)$ under policy ${\bf \pi}=\left\{{\pi}_1,\cdots,{\pi}_M\right\}$, the {action}-value function, i.e., the expected return, is given by
\begin{equation}
\label{q_value}
  Q^{\bf{\pi}}\left({\bf S}(t),{\bf a}(t)\right)= \mathbb{E}\left[\sum_{k=0}^{\infty}{\gamma}^k r_m(t+k)\bigg|{\bf S}(t),{\bf a}(t)\right]\,.
\end{equation}
With the Bellman equation, $Q^{\bf \pi}\left({\bf S}(t),{\bf a}(t)\right)$ is updated by
\begin{equation}
\label{q_value_update}
  Q^{\bf \pi}\!\left({\bf S}(t),{\bf a}(t)\!\right)\!\!=\! \!\mathbb{E}\big\{r_m(t)\!\!+ \!\!
  \gamma\mathbb{E}\!\left[Q^{\bf \pi}\!\left({\bf S}\!\left(t\!\!+\!\!1\right),{\bf a}(t\!\!+\!\!1\!)\big)\!\right]\!\right\}.
\end{equation}
The centralized training requires each agent to upload their experience into an experience replay buffer and download a batch of samples containing the experience of all agents in every time slot. However, frequent sample transmissions would entail significant communication overhead. 

{We improve the MADDPG by proposing semi-decentralized training with decentralized execution.
As shown in Fig.~\ref{fig:maddpg}, 
robot $m$ has an actor network, a target actor network, and a local replay buffer. The BS has a critic network, a target critic network, and a global experience replay buffer.  During this semi-decentralized training, only the critic networks at the BS are trained based on the global experience, while each robot can train its actor networks locally {based on the {gradients of the critic sent by the server instead of uploading its experiences to the server instantaneously. The robots can upload their experiences when wireless resources become abundant, under much less stringent delay requirements. Then, the server can update its replay buffer and train the critic network accordingly.}} }

{Next, we utilize the designed semi-decentralized method for solving the schedule and resource allocation of the D2D transmissions among the robots, which is modeled as a Dec-POMDP with error estimation, aiming to minimize the proposed time-average LoIU of a robot team.}
For conciseness, the observation $o_m(t)$, {action} $a_m(t)$, reward $r_m(t)$, and observation $o_m(t+1)$ are denoted as $o_m$, $a_m$, $r_m$ and $o'_m$. At every time slot, robot $m$ stores experience $(o_m,a_m,r_m,o'_m)$ into its local experience replay buffer ${\cal B}_m$ with a finite size of ${W}_{\cal B}$. The robot intermittently uploads its experience that has not been uploaded into a global replay buffer ${\cal B}$ at the BS, when the channel is in good condition. The global replay buffer has the size of ${W}_{\cal B}$. Once the replay buffer is full, the oldest sample is removed. 

The parameters $\theta_{m}^\mu$ and $\theta^{Q}$ of the actor and critic networks are updated per step at robot $m$, $\forall m$, and the BS, respectively. The parameters $\theta_{m}^{\mu'}$ of the target actor network and $\theta^{Q'}$ of the target critic network are updated by refreshing the actor and critic networks at robot $m$ and the BS, respectively, following the soft update technique, i.e.,
\begin{equation}
       \theta^{Q^{\prime}}=v \theta^{Q}+(1-v) \theta^{Q^{\prime}} \,; 
 \theta_{m}^{u^{\prime}}=v \theta_{m}^{u}+(1-v) \theta_{m}^{u^{\prime}} \,, \label{munetwork}
\end{equation}
where $v$ is a configurable soft update parameter.

\subsubsection{Update of Critic Network}
As illustrated in Fig.~\ref{fig:maddpg}, the BS pulls a mini-batch ${\cal K}$ that consists of $K$ samples $(\boldsymbol{o},\boldsymbol{a},\boldsymbol{r},\boldsymbol{o}^{\prime})$ from the central replay buffer ${\cal B}$ for the training of the critic network. Here, $\boldsymbol{o}=\left\{o_1,\cdots,o_M\right\}$, $\boldsymbol{a}=\left\{a_1,\cdots,a_M\right\}$, $\boldsymbol{r}=\left\{r_1,\cdots,r_M\right\}$, and $\boldsymbol{o'}=\left\{o'_1,\cdots,o'_M\right\}$. Correspondingly, robot $m$ pulls a mini-batch ${\cal K}_m$ that consists of $K$ samples $(o_m,a_m,r_m,o'_m)$ from its local replay buffer ${\cal B}_m$ to train its actor network. 
 
 For a sample $(o_m,a_m,r_m,o'_m) \in {\cal K}_m$, the target actor network of robot $m$ generates an {action} ${a}_{m}^{\prime}$ without exploration, based on the observation $o_m$. By collecting all {action}s based on ${\cal K}_m$, robot $m$ transmits $\boldsymbol{A}_m^{\prime}$ to the BS, i.e., 
\begin{equation}
    \label{A'}
    \boldsymbol{A}_{m}^{\prime}=\left\{a_{m}^{\prime}=\mu_{\theta_{m}^{{\mu}^{\prime} }}\left(o_{m}^{\prime}\right), \forall  o_{m}^{\prime} \in {\cal K}_m \right\} \in  \mathbb{R}^{K\times 2} \,. \\
\end{equation}
 The target critic network of the BS outputs a Q-value by inputting $\boldsymbol{o}^{\prime}$ 
 from each sample $(\boldsymbol{o},\boldsymbol{a},\boldsymbol{r},\boldsymbol{o}^{\prime})$, and {$\boldsymbol{a}^{\prime}=\{a_{1}^{\prime},\cdots,a_{M}^{\prime}\}$} received from the robots. 
 
 According to (\ref{globalreward}), the global reward, denoted as $r$, is equal to the reward of any robot, i.e., $r=r_m, \forall m \in {\cal M}$. Then, the Q-value can be evaluated of the BS at the critic network by minimizing its loss function, as given by 
\begin{equation}
    \label{lossFunction}
    \begin{aligned}
L\!\left(\theta^{Q}\right)\!\!=& \mathbb{E}_{\boldsymbol{o}, {\boldsymbol a}, {\boldsymbol r}, \boldsymbol{o^{\prime}}}\!\left[\left(r 
\!\!+\!\! \gamma 
Q^{\theta^{Q^{\prime}}}\!\!\left(\boldsymbol{o}^{\prime}, \boldsymbol{a}^{\prime} \right)\!\!
-\!\!Q^{\theta^{Q}}\!\left(\boldsymbol{o}, \boldsymbol{a} \right)\!\right)^{2}\right],
\end{aligned}
\end{equation}
where $Q^{\theta^{Q}}\left(\boldsymbol{o}, \boldsymbol{a}\right)$ is the {action}-value function with all observations $\boldsymbol{o}$ and all robots' {action}s $\boldsymbol{a}$ sampled from the central replay buffer as the inputs and the Q-value as the output. 
We use batch normalization to normalize the various observations for the stability of policy gradients in the critic network. 

\subsubsection{Update of Actor Network}
{The {action} space consists of discrete robot selection and resource {allocation}. In this paper, DDPG 
 is adapted to handle discrete {action}s by incorporating the Gumbel-Softmax (GS) technique \cite{jang2016categorical}. GS allows discrete {action}s to be approximated as continuous distributions during the forward pass, while enabling backpropagation through the discrete {action} space. A GS sampler produces differentiable samples during training, which computes the gradients of an {action} sample with respect to the parameters at the output layer of the actor network. 
With the sampled observations} $\left\{{o}_m, \forall o_m \in {\cal K}_m\right\}$, the actor network of robot $m$ generates and uploads an {action} set $\boldsymbol{a}_{m}^{\mu}$ to the critic network: 
\begin{equation}
\label{A}
    \boldsymbol{a}_{m}^{\mu}=\left\{a_{m}^{\mu}=\mu_{\theta_{m}^{\mu}}\left(o_{m}\right), \forall o_{m} \in {\cal K}_m\right\} \in \mathbb{R}^{K \times 2} .
\end{equation}

 {For each robot $m$, the critic network of the BS collects $K$ gradients ${\cal G}_m$ of its {action}s, where each gradient is obtained based on $a_{m}^{\mu} \in \boldsymbol{a}_{m}^{\mu} $, observations $\boldsymbol{o}$, and {action}s $\boldsymbol{a}$ (except $a_m$), in each sample $(\boldsymbol{o},\boldsymbol{a},\boldsymbol{r},\boldsymbol{o}^{\prime})$ from $\mathcal{K}$, as given by}
\begin{subequations}
\label{Gradients}
    \begin{align}
    \label{Gradient}
        &{\cal G}_m=\left\{{\cal G}_m(\boldsymbol{o}),\forall \boldsymbol{o} \in {\cal K} \right\}\in \mathbb{R}^{K \times 2} \,;  \\
        &
        \begin{aligned}
            {\cal G}_m(\boldsymbol{o})=&\nabla_{a_{m}} Q^{\theta^Q}\left({\boldsymbol{o}}, a_{1}, \cdots, \left. a_{M}\right)\right|_{a_{m}=a_{m}^{\mu}}\in \mathbb{R}^{2} \,.
        \end{aligned}
    \end{align}
\end{subequations}
 {The BS sends the gradients ${\cal G}_m$ to robot $m$. ${\cal G}_m$ can differ from ${\cal G}_n$, if $m \neq n$.} Upon receiving ${\cal G}_m$, the actor network of robot $m$ can be updated locally using the policy gradient, i.e.,
\begin{equation}
    \label{actorUpdate}
        \nabla_{\theta_{m}^{\mu}} J\left(\mu_{\theta_{m}^{\mu}}\right)=\mathbb{E}_{\boldsymbol{o}}\left[\nabla_{\theta_{m}^{\mu}} \mu_{\theta_{m}^{\mu}}\left(o_{m}\right) {\cal G}_m(\boldsymbol{o}) \right]  \,.
\end{equation}

\textbf{Algorithm \ref{algorithm}} summarizes the proposed semi-decentralized MADDPG, where the robots do not need to download samples for training their actors, and they can upload their local experiences with high rewards upon the request of the BS. This contributes to the reduced communication overhead and increased flexibility. Let $\beta_{\mathrm{o}}$, $\beta_{\mathrm{a}}$, and $\beta_{\mathrm{r}}$ denote the sizes of $o_m$, $a_m$, and $r_m$ ($\forall m \in \cal{M}$), 
respectively; and $\beta_{\mathrm{p}}$ denote the data size of ${\cal G}_m(\boldsymbol{o})$ ($\forall m \in \cal{M}$).
At slot $t$, a robot uploads its {action} set $\boldsymbol{A}_{m}^{\prime}$ in (\ref{A'}) and $\boldsymbol{a}_{m}^{\mu}$ in (\ref{A}), and downloads the gradients ${\cal G}_m$ in~(\ref{Gradient}). As a result, the uploaded and downloaded data sizes are $2KM\beta_{\mathrm{a}}$ and $KM\beta_{\mathrm{p}}$, 
respectively.
By contrast, 
in the traditional MADDPG implementation with localized actor and critic networks at each robot and a shared experience replay buffer at the BS, each robot needs to upload its local experience with the size of $M(2\beta_{\mathrm{o}}\!+\!\beta_{\mathrm{a}}\!+\!\beta_{\mathrm{r}})$ in addition to its {action} set $\boldsymbol{A}_{m}^{\prime}$, and download a mini-batch of 
$K$ global samples with the size of $KM^2(2\beta_{\mathrm{o}}\!+\!\beta_{\mathrm{a}}\!+\!\beta_{\mathrm{r}})$ at slot~$t$. Thus, the overall uploaded and downloaded data sizes of traditional MADDPG are $KM\beta_{\mathrm{a}}\!+\!M(2\beta_{\mathrm{o}}\!+\!\beta_{\mathrm{a}}\!\!+\!\!\beta_{\mathrm{r}})$ and $KM^2(2\beta_{\mathrm{o}}\!+\!2\beta_{\mathrm{a}}\!+\!\beta_{\mathrm{r}})$, respectively. 

{
The proposed semi-decentralized MADDPG framework removes the need for each robot to transmit its observations instantaneously to the server. This avoids the potential air interface congestion that could arise in a centralized control approach, e.g., the centralized implementation of DDPG/MADDPG for collaborator selection and resource allocation.
Under the semi-decentralized MADDPG architecture, part of the required computation (i.e., the training of the critic networks) is offloaded to the edge. This reduces the computational burden on the robots, balancing the load between mobile robots and edge servers.
Additionally, the MADDPG model can be trained {offline} using historical data and applied online for inference, enabling computationally intensive operations intermittently without putting excessive strain on the robots and server in real-time.}

	\begin{algorithm}[t]
	\caption{Proposed semi-decentralized MADDPG}
        \label{algorithm}
	\begin{algorithmic}[1]
	\REQUIRE $M$, $\theta_{m}^{u}$, $\theta_{m}^{u^{\prime}}$, $\theta^{Q}$, $\theta^{Q^{\prime}}$, $\mathcal{B}$, $W_{\cal B}$, ${\cal B}_m$, $K$, $\gamma$, $v$.
	\ENSURE $\theta_{m}^{u}$, $\forall m \in \mathcal{M}$.
	\FOR {episode = $1, 2, \cdots, L_{e}$}
	\STATE Obtain the original observation $\boldsymbol{o}$ by (\ref{agentObeservation}).
	\FOR {slot = $1, 2, \cdots, L_{s}$}
        \STATE $//$ Decentralized execution
        \FOR {robot $m = 1, 2, \cdots, M$} 
        \STATE The actor network of robot $m$  chooses its {action} $a_{m}=\mu_{\theta_{m}^{\mu}}(o_{m}) $. The {action} is performed, obtaining reward $r_m$ and the new observation $o'_{m}$.
 	\IF {the local replay buffer $\mathcal{B}_m<W_{\cal B}$}
	\STATE Record the transition $(o_m,a_m,r_m,o'_m)$ into $\mathcal{B}_m$.
	\ELSE 
	\STATE Replace the most outdated sample in $\mathcal{B}_m$ with the current transition $(o_m,a_m,r_m,o'_m)$.
	\ENDIF
        \IF {the BS requests experience to be uploaded}
        \STATE Upload its experience to the BS for updating $\mathcal{B}$ 
        \ENDIF
	\ENDFOR
        \STATE $//$ Semi-decentralized training
        \STATE   {Robot $m \in \cal{M}$ samples a mini-batch of $K$ samples from $\mathcal{B}_m$. With the mini-batch, $\boldsymbol{A}_{m}^{\prime}$ and $\boldsymbol{a}_{m}^{\mu}$ are obtained and uploaded to the BS from its target actor and actor, respectively.}
        \STATE   {The BS samples a mini-batch of $K$ samples from $\mathcal{B}$. Its critic obtains ${\cal G}_m$  in (\ref{Gradients}) and sends it to robot $m$.
        The BS also updates its target critic and critic using  (\ref{munetwork}) and (\ref{lossFunction}), respectively.}
 	\STATE   {Robot $ m \in \cal{M}$ updates its target actor and actor based on (\ref{munetwork}) and (\ref{actorUpdate}), respectively.}
	\ENDFOR
	\ENDFOR
	\end{algorithmic}
	\end{algorithm}

 \subsection{  {Convergence Analysis}}
We analyze the convergence of the proposed MADDPG framework by proving the existence of the Nash Equilibrium (NE) in the framework. This starts by interpreting the framework as a distributed resource scheduling game $\Gamma=({\cal M}, {\cal A}, {\cal U})$ among the robots. The game consists of:
\begin{itemize}
\item The set of players ${\cal M} = \{1,2,\cdots,M\} $;
\item The {action} spaces of all robots ${\cal A}=\left\{{\cal A}_m\right\}_{m\in {\cal M}}$; and
\item The set of the joint costs of all robots ${\cal U}=\left\{{\cal U}_m\right\}_{m\in {\cal M}}$, 
\end{itemize}
where ${\cal U}_m$ is the immediate reward of robot $m$, i.e.,
\begin{equation}
\label{joint cost m}
      {\cal U}_m=r_m  \,. \\
\end{equation}

According to (\ref{optimizationP}) and (\ref{joint cost m}), {robot $m$ wishes to maximize its reward during the game:}
\begin{equation}
\label{purpose}
    {\max}_{a_m \in {\cal A}_m}  {r}_m\left(a_m,{\bf a}_{-m}\right)    \,, \\
\end{equation}
where ${\bf a}_{-m}=\{a_1,\cdots,a_{m-1},a_{m+1},\cdots,a_M\}$ collects the {action}s of the robots other than robot $m$.


\textit{Definition 1 (Nash Equilibrium): For game $\Gamma$, we call ${\bf a}^\ast=(a_1^\ast,a_2^\ast,\cdots,a_M^\ast)$ a Nash Equilibrium if and only if no robot can further improve its {reward} by unilaterally changing its {action} at the equilibrium ${\bf a}^\ast$, i.e.,}
\begin{equation}
\label{NE}
      {r}_m\left(a_m^\ast,{\bf a}_{-m}^\ast\right){>}{r}_m\left(a_m,{\bf a}_{-m}^\ast\right)  \,. \\
\end{equation}

\textit{Definition 2 (Potential Game): A game is a potential game if there exists a potential function $\Phi$: ${\bf a}=(a_1,a_2,\cdots,a_M) \to \Bbb{R}$ such that ${\forall m} \in {\cal M}$ :}
\begin{equation}
\label{PG}
      {r}_m\!\left(a_m,\!{\bf a}_{-\!m}\right)\!-\!{r}_m\!\left(a'_m,\!{\bf a}_{-\!m}\right)\!\!=\!\! 
      \Phi\!\left(a_m,\!{\bf a}_{-\!m}\right)\!-\!\Phi\!\left(a'_m,\!{\bf a}_{-\!m}\right). 
\end{equation}

\textit{Theorem 1: $\Gamma$ is a potential game and has an NE with a finite number of unilateral overhead updates.}

\textit{Proof:} By Definition 2, we define a function $\Phi$ : ${\bf a}=(a_1,a_2,\cdots,a_M) \to \Bbb{R}$ as
\begin{equation}
\label{PoF}
      {\Phi}\left(a_m,{\bf a}_{-m}\right)= \frac{1}{M} \sum_{m'=1}^{M} {r}_{m'}(a_{m'},{\bf a}_{-m'}) \,. \\
\end{equation}
Consider the semi-decentralized training process illustrated in Fig.~\ref{fig:maddpg}. Each robot takes the global reward as its reward. The reward is evaluated centrally by the critic network at the BS. According to (\ref{globalreward}) and (\ref{joint cost m}), the reward ${r}_{m}(a_{m},{\bf a}_{-m})$ is 
\begin{equation}
\begin{aligned}
\label{pro1}
      {r}_{m}(a_{m},{\bf a}_{-m})= & \frac{1}{M} \sum_{m'=1}^{M} r_{m'}^{\mathrm{l}} \,, \\
\end{aligned}
\end{equation}
which is also the global reward and does not change over $m$.

Then, the potential function ${\Phi}\left(a_m,{\bf a}_{-m}\right)$ is given by 
\begin{equation}
\begin{aligned}
\label{PoF2}
      {\Phi}\left(a_m,{\bf a}_{-m}\right)= & \frac{1}{M} \sum_{m'=1}^{M} \left( \frac{1}{M} \sum_{m'=1}^{M} r_{m'}^{\mathrm{l}} \right) \\
      = & \frac{1}{M} \cdot M {r}_{m}(a_{m},{\bf a}_{-m}) 
      = {r}_{m}(a_{m},{\bf a}_{-m}). 
\end{aligned}
\end{equation}
Consequently, (\ref{PG}) is obtained. As a potential game, $\Gamma$ surely has an NE.

{
Consider the game is a one-step {decision} process; i.e., $Q^{\bf \pi}\!({\bf S}(t+1),{\bf a}(t+1))$ is independent with $Q^{\bf \pi}\!({\bf S}(t),{\bf a}(t)\!)$. Thus, the second term on the right-hand side of \eqref{q_value_update} is $0$ and, subsequently, $Q^{\bf \pi}\!\left({\bf S}(t),{\bf a}(t)\!\right)\!\!=\! \!\mathbb{E}\{r_m(t)\}$~\cite{geng2024balancing}. From \eqref{PoF2}, it follows that $\nabla_{a_{m}} Q^{\theta^Q}({\boldsymbol{o}}, {\boldsymbol{a}})=\nabla_{a_{m}} {\Phi}(a_m,{\bf a}_{-m})$, i.e., each agent’s local gradient step aligns with $\nabla_{a_{m}} {\Phi}(a_m,{\bf a}_{-m})$, which implies $\Phi(\mathbf{a})$ is non-decreasing. 

Assume that the actor and critic networks have bounded parameters and Lipschitz gradients. By designing the learning rates satisfying standard stochastic approximation conditions~\cite{robbins1951stochastic} and exploration noise diminishing over time, the actor-critic updates converge almost surely to a stationary point of the expected update dynamics~\cite{konda2000actor}, i.e.,
\begin{equation}
  \lim_{t \to \infty} \theta_m^{\mu} \;\to\; (\theta_m^{\mu})^*,
  \quad
  \text{where } \nabla_{\theta_m^{\mu}} J_m((\theta_m^{\mu})^*) = 0.
\end{equation}

Based on \eqref{actorUpdate}, any stationary point of the global updates must satisfy $\nabla_{\mathbf{a}} \Phi(\mathbf{a}^*) = \mathbf{0}$, with $\mathbf{a}_{m}^*=\mu_{(\theta_m^{\mu})^*}$.
According to \textit{Definitions 1} and \textit{2}, the policy converges to a local maximum of $\Phi(\mathbf{a})$, which is a local NE of the potential game.
In this way, the convergence of the MADDPG framework is guaranteed.
In a more general or complex setting, the convergence properties can be empirically validated, as in Section~V-C.
}

\subsection{  {Complexity Analysis}}
The structures of the actor networks and critic networks determine the computational complexity of the proposed algorithm. The actor network of each robot contains $I_{\mathrm{a}}$ fully connected layers with $n_i^{\mathrm{a}}$  neurons in the $i$-th layer. The critic network at the BS contains $J_{\mathrm{c}}$ fully connected layers with $n_j^{\mathrm{c}}$ neurons in the $j$-th layer. The computational complexities of an actor network and a critic network are ${\cal O}(\sum\nolimits_{i = 1}^{I_{\mathrm{a}}} {n_{i - 1}^{\mathrm{a}}n_i^{\mathrm{a}}} )$ and ${\cal O}(\sum\nolimits_{j = 1}^{J_{\mathrm{c}}} {n_{i - 1}^{\mathrm{c}}n_i^{\mathrm{c}}} )$, respectively. Hence, the computational complexity of training is {${\cal O}({N_{\mathrm{e}}}{N_{\mathrm{s}}}{K}(M\sum\nolimits_{i = 1}^{I_{\mathrm{a}}} {n_{i - 1}^{\mathrm{a}}n_i^{\mathrm{a}}}  + \sum\nolimits_{j = 1}^{J_{\mathrm{c}}} {n_{j - 1}^{\mathrm{c}}n_j^{\mathrm{c}}}))$}. Here, $N_{\mathrm{e}}$ is the number of episodes. $N_{\mathrm{s}}$ is the maximum number of training steps per episode. In contrast, the complexity of online execution is only ${\cal O}(\sum\nolimits_{i = 1}^{I_{\mathrm{a}}} {n_{i - 1}^{\mathrm{a}}n_i^{\mathrm{a}}} )$ per robot, and only occurs at the actor network of a robot.

	\section{Simulation Results}

 

\begin{table}[t]
\caption{Simulation parameter configuration}
\label{parameters}
\centering
\begin{tabular}{|*{2}{l|}}
\hline
\textbf{Parameter} & \textbf{Value} \\
\hline
Cell radius & 150 m \\
D2D pair distance & $\textless$40 m \\
RB bandwidth $W$& 180 kHz \\
BS transmit power $P_b$& 46 dBm \\
Robot transmit power $P_r$& 13 dBm \\
Noise spectral density & -114 dBm/Hz \\
Cellular link path loss & $128.1+37.6\log_{10}{\left(d\right)} $ \\
Pathloss constant & 1 \\
Pathloss exponent & 4 \\
 {Robot SINR threshold $\varsigma_{th}$ \cite{li2019multi}}& 0 dB \\
Duration of a time slot & 10 ms \\
Replay buffer size $W_{\cal B}$& $10^6$  \\
Batch size $K$& 1,000  \\
Initial exploration probability & 0.1  \\
Discount factor $\gamma$& 0.97  \\
Learning rate & 0.01  \\
Soft update parameter $v$& 0.001  \\
penalty coefficient $R^s$& 1 \\
\hline
\end{tabular}
\end{table}

\subsection{Simulation Settings}
{The simulations were conducted using a custom-designed network simulation environment implemented in Python, which integrates PyTorch for DRL and communication modeling.}
Consider a team of automated guided vehicles (AGVs) within the coverage of a BS with a radius of 150 m \cite{sun2017emm}. The robots are uniformly and randomly distributed within the coverage, each having a communication range of up to 40 m. Each robot can randomly move up to 0.2 m within a slot of 10 ms, i.e., 20 m/s, in a randomly selected direction. The standard deviation $\sigma_{n}$ of changes in the estimation errors of the {status} $x_{n}^t$ is uniformly and randomly taken from $[0.001,10]$ per slot.
The upper limit of the estimation errors is 
{
$E_{n,m}=z_{n,m}\sigma_{n}^2$, where by default, $z_{n,m}$ and ${{\cal D}_m^t}$ are randomly and uniformly taken from $[0.2,15]$ and $[2, 100]$ ms, respectively.} The sizes of {status} information, $\alpha_{n,m}(t)$, $\forall m,n \in \cal{M}$, differ among different robots, and remain unchanged over time. They are uniformly and randomly taken from $[40, 120]$ bytes. 
{Moreover, we consider varying levels of resource availability and robot density by setting $M=5$ to 25 robots and $J=2$ to 6 RBs, each RB having a bandwidth of 180~kHz.}


  {Our simulations were performed using an NVIDIA GeForce GTX 1080 Ti GPU.} Using the PyTorch framework, the proposed semi-decentralized MADDPG framework consists of two deep neural networks (DNNs) for each robot (i.e., an actor network, and a target actor network) and two DNNs for the BS (i.e., a critic network, and a target critic network).
The actor network comprises an input layer, two hidden layers (128 and 256 neurons), and an output layer. The ReLU activation function is applied to the hidden layers, while the softmax activation function is used at the output layer.
For the critic network, the numbers of neurons are 256, 128, and 64 in the three fully-connected hidden layers. The ReLU activation function is employed for the hidden layers. The output layer performs linear transformation to produce the Q-values for the {state}-action pairs.
The simulation parameters can be found in Table~\ref{parameters} unless specified otherwise.

	\begin{figure}[t]
	\centering
	\includegraphics[width=0.45\textwidth]{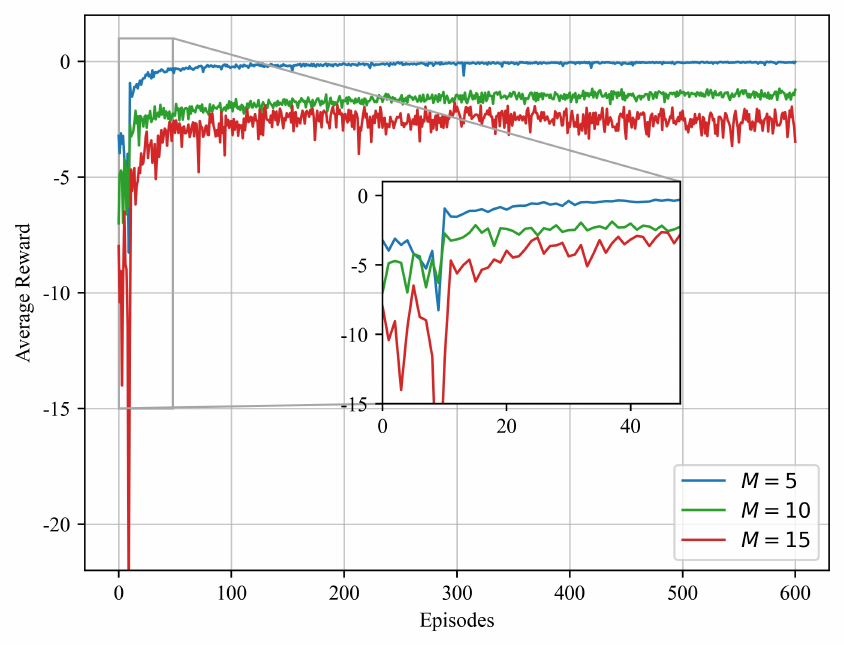}
	\caption{The  {convergence performance of the semi-decentralized resource scheduling scheme, where $M=5, 10, 15$, $J=4$ and $\left|{\cal C}_m \right|=4$.}}
	\label{training}
	\end{figure}

 \subsection{Benchmarks and Metrics}
 For comparison, four benchmark schemes are considered: 
 \begin{itemize}
     \item DDPG: Each robot is an agent to independently perform the DDPG algorithm for learning its policy, including collaborator selection and resource allocation \cite{kwon2020multiagent}. Both actor and critic networks reside in a robot.
    \item DQN: Each robot is an agent to perform the DQN algorithm for learning independently its policy~\cite{li2019multi}. 
    \item {Advantage Actor-Critic (A2C): Each robot is an agent to independently run the A2C algorithm for learning its policy~\cite{mnih2016asynchronous}.
     Both critic and actor networks reside in a robot to compute advantages and update policy.}
    \item {Soft Actor-Critic (SAC): Each robot is an agent that performs SAC to learn its policy independently~\cite{haarnoja2018soft}.
     In each robot, an actor and two Q-value networks are used to optimize the policy and value function.}
        \item {Time Division Multiplexing (TDM): Each time slot is evenly divided into discrete mini-slots, and each D2D pair is assigned a specific mini-slot for transmission~\cite{vaezi2019multiple}. }
     \item Random policy: {Each robot randomly selects its collaborators and RBs for its D2D pairs (i.e., $l_{n,m}^{j,t}$), which are uniformly and randomly taken from all possible {action}s from its {action} space.} 
     \item {All-allocated policy: Unlike this random policy, all transmitting robots are allowed to transmit using randomly allocated wireless resources {(i.e., $\sum_{j=1}^{J} l_{mn}^{j,t}=1$)}.}
    \item {Threshold-based policy: If LoIU exceeds a specific threshold (i.e., 
    $\zeta <\mathcal{F}_{m}(t)\leq 1 $), the all-allocated policy is employed for its collaborators. For the other metrics (i.e., AoI, AoII, UoI, and AoI), if the metric of a collaborator's {status} is larger than the given threshold (i.e., $\zeta\,{\cal D}_{m}^{t}$ for AoI, AoII, and AoCI, and $\zeta\, E_{nm}$ for UoI), an RB is randomly allocated to the collaborator.}
 \end{itemize}
 We compare the proposed metric, i.e., LoIU, with the existing metrics, i.e., AoI\cite{kaul2012real}, AoII\cite{maatouk2020age},  UoI\cite{zheng2020urgency}, and AoCI~\cite{wang2021age}.   

\subsection{Convergence Analysis}
	Fig.~\ref{training} demonstrates that the proposed algorithm achieves stable convergence within $100$ episodes. This can be attributed to a semi-decentralized training and decentralized execution framework. During the semi-decentralized training, the BS utilizes samples drawn from the global replay buffer for updating the critic network; each robot draws samples from local replay buffer and receives the parameters from the BS to update its actors. As a result, the algorithm addresses the non-stationary environment changes encountered by the traditional single-agent RL. 
 The algorithm can attain higher system rewards by capturing the collaboration and requirements of the robots in the reward function (\ref{localReward}) and (\ref{globalreward}).


    \begin{figure}[t]
    \centering
    \includegraphics[width=0.45\textwidth]{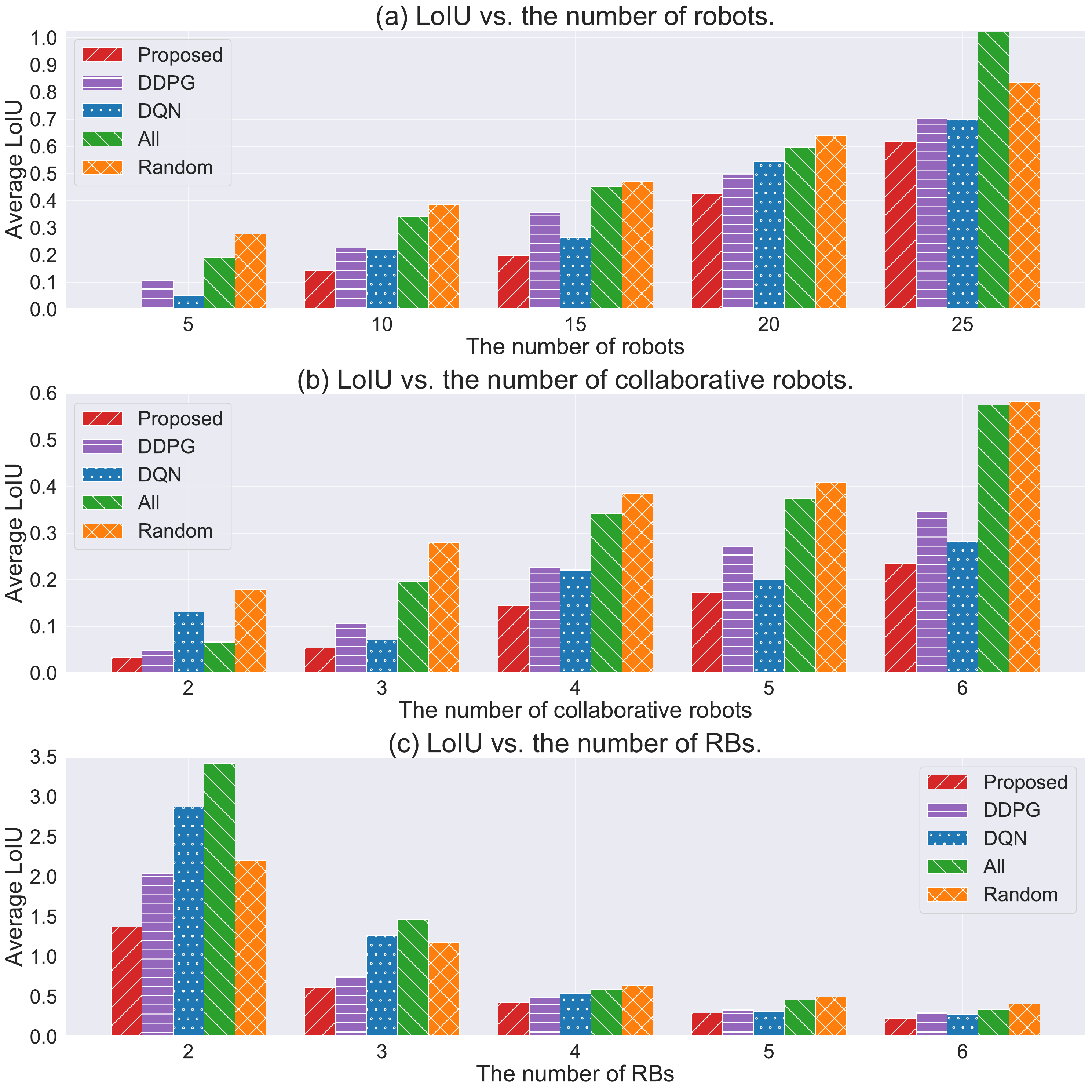}
    \caption{Average LoIU vs. the number of robots ( $\left|{\cal C}_m \right|=4$, $J=4$), {collaborative} robots ($M=10$, $J=4$), and RBs ($M=20$, $\left|{\cal C}_m \right|=4$).}
    \label{LoIU_ro}
\end{figure}



 \subsection{LoIU under Different Strategies}
	 {Figs.~\ref{LoIU_ro}(a) and~\ref{LoIU_ro}(b)} show the changes of LoIU with the increase of robots and their collaborators, respectively. The LoIU increases under all methods because limited wireless resources and interference constrain transmissions among the robot team. 
     {
     When there are 5 robots, the average LoIU of the proposed method is very small, approximately
$0.00506667$. This is because when resources are relatively abundant, the method is effective in optimizing resource allocation, leading to negligible information utility loss.} 
As the robots and collaborators increase, wireless resources become increasingly constrained to support the freshness and correctness of the {operational status} information. More {status} updates are overdue and violate the bounds of estimation errors, increasing LoIU.

{When the robots operate in close proximity, both resource contention and interference between the robots inevitably increase, which can lead to performance degradation. Nevertheless, our method can more effectively mitigate the impact of increased interference compared to the benchmarks in Fig.~\ref{LoIU_ro}(a), keeping LoIU at a relatively low level for efficient and timely information exchange for collaborative tasks. This is because our method is able to adaptively adjust the transmission from collaborators and the selection of RBs.}

LoIU is the lowest under the proposed algorithm,
since the random and all-allocated policies do not consider the collaboration and competition of robots. 
This leads to more delays and distortions of {the operational status} updates, especially with many robots. LoIU is less under the single-agent DDPG and DQN than the other two benchmarks, as a robot can dynamically update the policy by maximizing LoIU for the robots in its range. However, the single-agent DDPG and DQN are limited to local observations. 

{As depicted in  {Fig.~\ref{LoIU_ro}(b)}, as the number of collaborators increases, the average LoIU of the robot team increases. This is due to tighter resource availability and increased interference, which hinder efficient information transmission.} Fig.~\ref{LoIU_ro}(c) indicates that the proposed algorithm outperforms the four benchmarks in the network with limited RBs. When RBs are abundant, LoIU is still lower under our algorithm than the benchmarks because the algorithm fully considers the global environment and adjusts the policy to minimize LoIU.

	\begin{figure}[t]
	\centering
	\includegraphics[width=0.45\textwidth]{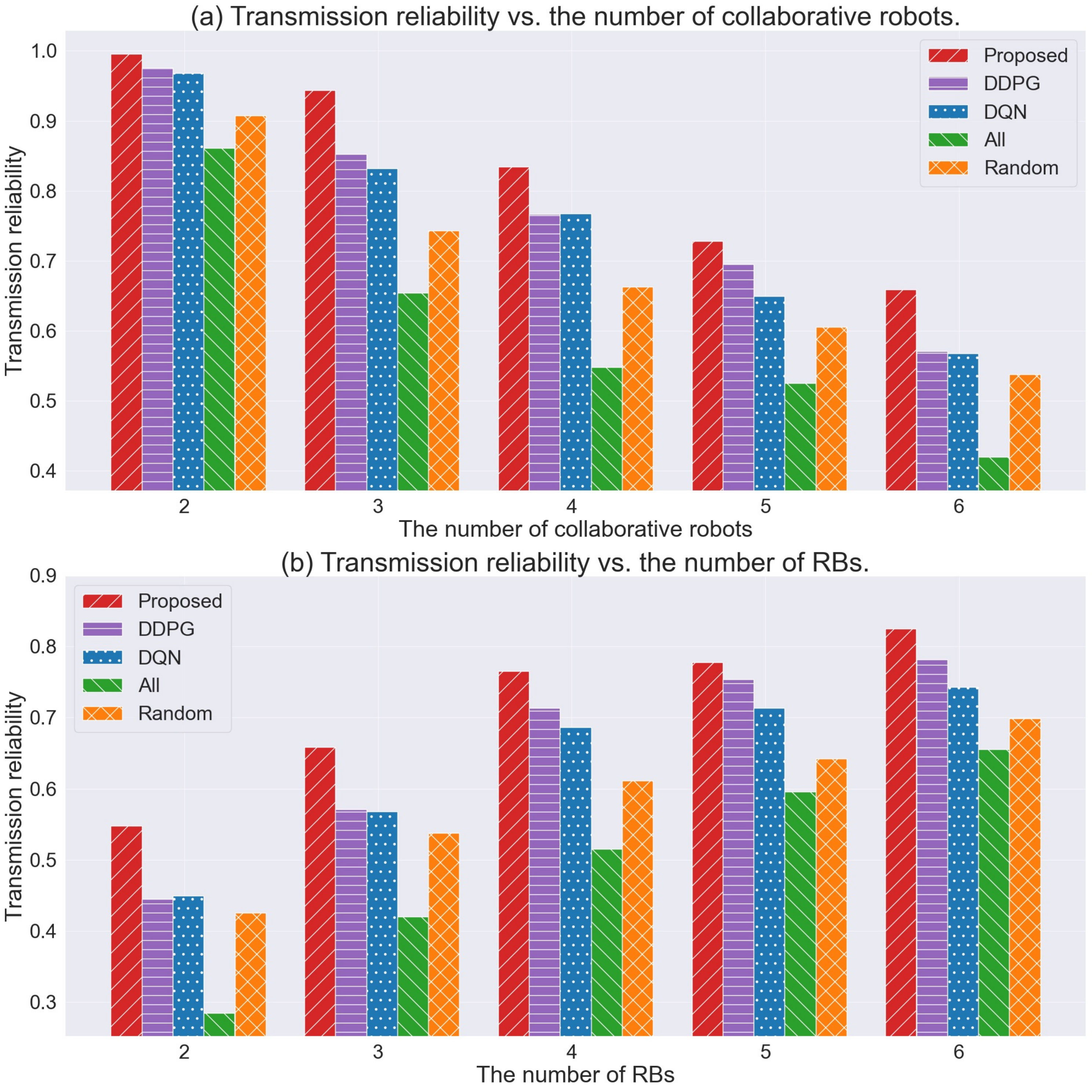}
	\caption{{The transmission reliability of the proposed method vs. the number of collaborative robots ($M=10$, $J=3$), and RBs ($M=10$, $\left|{\cal C}_m \right|=6$).}}
	\label{trans_co}
	\end{figure}


 \subsection{Transmission Reliability under Different Strategies}
	We evaluate the transmission reliability of the proposed algorithm by considering the transmission latency and the interference of the D2D links. If a transmission satisfies Constraint \textbf{C1} with an SINR that is higher than the minimum SINR (i.e., $\rho_{n,m}^t$=1), then the transmission is regarded as successful. The transmission reliability is measured by the average ratio of successful transmissions~\cite{cui2022multi}. 
 
  {Fig.~\ref{trans_co}(a)} shows the transmission reliability declines as the number of collaborative robots (i.e., $\left|{\cal C}_m \right|$) grows from 2 to 6. 
  The shortage of wireless resources causes severe interference between the D2D links as the D2D pairs grow.
  With each robot having six collaborators, our method improves transmission reliability by 15.38\% and 56.99\%, over the DDPG-based and all-allocated policies. The method captures robot cooperation and competition through semi-decentralized training. \textbf{C1} is fulfilled by properly designing a penalty function in the reward, improving transmission reliability.  
 {Fig.~\ref{trans_co}(b)} shows that the transmission reliability of all schemes increases as the number of available RBs grows. The average ratio of the proposed algorithm surpasses those of the four benchmarks, particularly in scenarios with limited available wireless resources. The reason is that the algorithm accounts for interference between links and real-time requirements of {the operational status} information to utilize resources efficiently.

\subsection{Task Reliability under Different Metrics}

	\begin{figure}[t]
	\centering
	\includegraphics[width=0.45\textwidth]{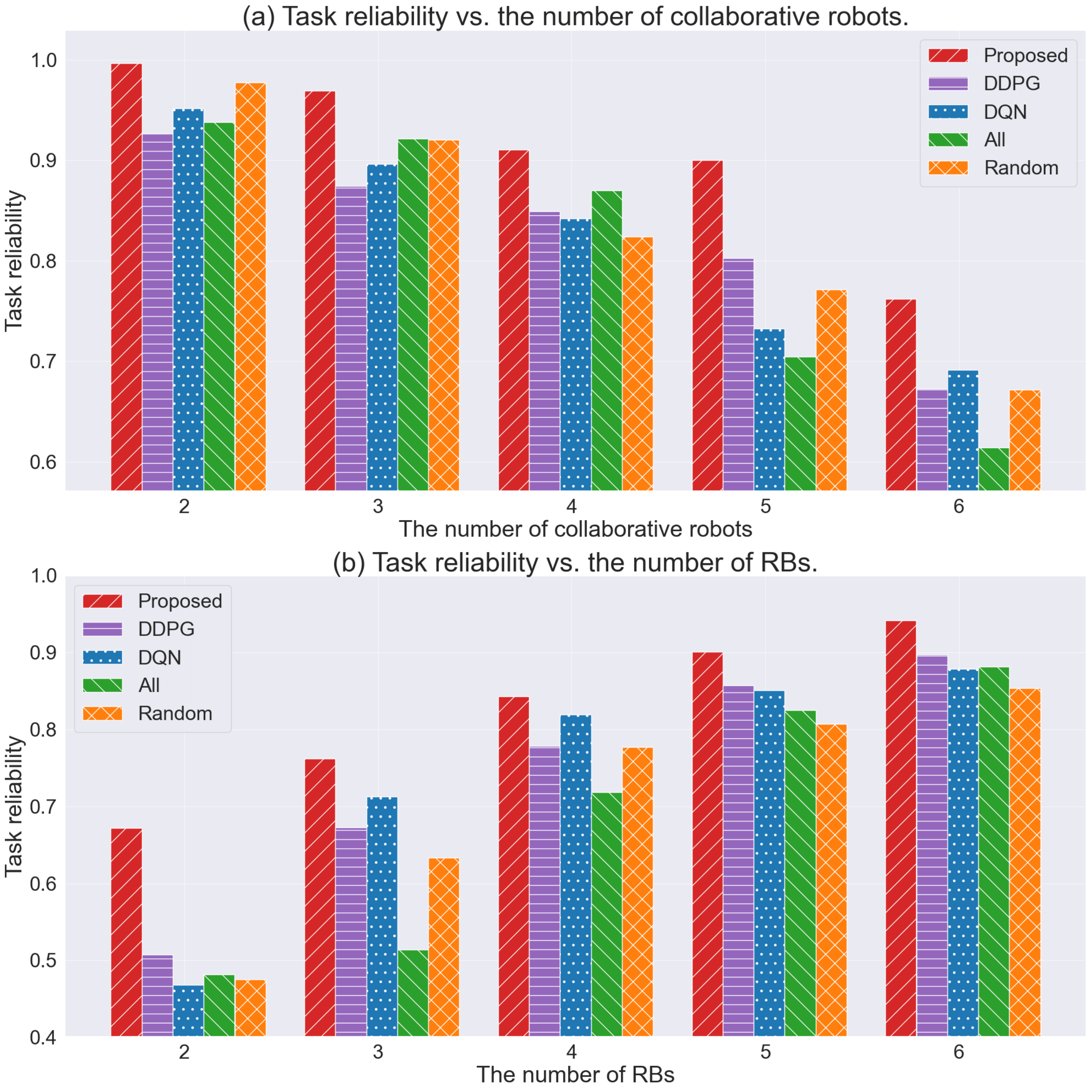}
	\caption{{The task reliability of the semi-decentralized scheduling scheme vs. the number of robots ($M=10$, $J=3$), and RBs ($M=10$, $\left|{\cal C}_m \right|=6$).}}
	\label{task_co}
	\end{figure}

 
We also compare LoIU with existing metrics {(i.e., AoI, AoII, UoI, and AoCI)}. 
 Considering a scenario where a team of AGVs share their velocities with each other. If an AGV's perception of another's velocity deviates beyond a certain threshold and fails to update its accurate velocity for a certain duration, there is a risk of collision and the transportation task fails. 
{Therefore, we define \textit{task reliability} as 
 the proportion of instances where a robot (e.g., robot $m$) successfully receives accurate {operational status} updates from its collaborators (e.g., robot $n$, $n\neq m$) within their respective deadlines (i.e., ${\cal D}_m^t$) or accurately estimates the {status} updates within their respective estimation error thresholds (i.e., $E_{n,m}$) in a time slot $t$, ensuring both latency and accuracy thresholds are satisfied. 
 This definition of \textit{task reliability} aligns with the objectives of multi-robot collaboration, where effective coordination relies on accurate and fresh information exchange.}

 {Fig.~\ref{task_co}(a)} evaluates the task completion rate as the size of each collaborative robot set, i.e., $\left|{\cal C}_m \right|$, increases from $2$ to $6$. Task reliability declines under all considered schemes as the number of collaborative robots increases. LoIU contributes to stronger reliability than the other three metrics, e.g., when many robots require collaboration and information sharing. This is because LoIU captures the collaboration and heterogeneous information freshness and correctness requirements, instructing efficient collaborator selection and resource allocation.
 {Fig.~\ref{task_co}(b)} shows the completion rates of tasks based on LoIU and the other existing metrics rise, as the wireless resources become abundant. LoIU outperforms the other metrics. Even when wireless resources are scarce, the task completion rates under LoIU remain relatively high. The reason is that the LoIU-based policy efficiently utilizes wireless resources for {operational status} updating by considering the latency and accuracy requirements of different {status} information and the collaboration between the robots.

 \begin{table}
    \centering
    \caption{ Task reliability of different methods and metrics {($z_{n,m}\in[1, 5]$, ${{\cal D}_m^t}\in [2,100]$ milliseconds)}}
    \begin{tabular}{l|c|c|c|c|c}
     \hline
     & LoIU & AoI& AoII & UoI& AoCI \\
     \hline
       Proposed  & \textbf{0.759}  & 0.734  & 0.704  & 0.675  & 0.613 \\
       DDPG  & {0.717}  & \textbf{0.682}  & 0.620  & 0.664  & 0.528 \\
       DQN  & {0.718}  & 0.672  & 0.664  & 0.645  & 0.581 \\
       {A2C}  & {0.627}  & 0.602  & 0.602  & 0.619  & 0.602 \\
       {SAC}  & {0.612}  & 0.548  & 0.567  & 0.584  & 0.601 \\
    {TDM}  & {0.558}  & 0.584  & 0.490  & 0.596  & 0.599 \\
       {All}  & {0.523}  & 0.535  & 0.535  & 0.553  & 0.542 \\
       {Random}  & {0.551}  & 0.557  & 0.560  & 0.575  & 0.559 \\
       {Threshold ($\zeta=0.1$)}  & {0.404}  & 0.216  & 0.216  & 0.215  & 0.216 \\
    {Threshold ($\zeta=0.5$)}  & {0.400}  & 0.216  & 0.216  & 0.215  & 0.216 \\
       \hline
    \end{tabular} %
    \label{table_methods}
\end{table}

Table \ref{table_methods} compares the different schemes and metrics in task reliability, with $M=15$, $J=4$, and $|{\mathcal{C}}_m|=4$. Under a given metric (e.g., LoIU, AoI, etc.), the proposed semi-decentralized MADDPG outperforms its benchmarks {(i.e., the DDPG, DQN, A2C, SAC, TDM, random, all-allocated, and threshold-based policies)}. Under the same semi-decentralized MADDPG scheme, the use of LoIU helps improve task reliability compared to the other metrics (i.e., AoI, AoII, UoI, and AoCI). 
{
Under the new semi-decentralized MADDPG and LoIU framework, task reliability exceeds that of the best existing approach (DDPG and AoI) by 11.29\%. This is because the DQN, A2C, and SAC policies allocate resources without considering the competition among agents and do not
evaluate the global performance, whereas the TDM, random, threshold-based, and all-allocated methods do not adapt to different requirements and communication {state}s. By contrast, the proposed scheme and metric consider the requirements of robots, time-varying {state}s, and the global objective, thus improving global performance.} 

{
\begin{table}
\tabcolsep=4pt
    \centering
    \caption{{Task reliability of different metrics and settings}}
    \begin{tabular}{c|c|c|c|c|c|c|c}
     \hline
     $z_{n,m}$ & ${{\cal D}_m^t}$ (ms) & LoIU & AoI & AoII & UoI & AoCI & Gain (\%) \\
     \hline
       $[0.2, 5]$& $[2,200]$ & \textbf{0.679}  & 0.625  & 0.623  & 0.623  & \textbf{0.643} & 5.60 \\
       $[0.2, 10]$ & $[2,200]$  & \textbf{0.734}  & 0.663  & 0.655  & 0.655  & \textbf{0.667} & 10.04 \\
        $[0.2, 15]$ & $[2,200]$ & \textbf{0.788}  & \textbf{0.713}  & 0.698  & 0.693  & 0.705 & 10.52 \\
        $[0.2, 15]$ & $[2,100]$  & \textbf{0.767}  & 0.689  & \textbf{0.695}  & 0.687  & \textbf{0.695} & 10.36 \\
        $[0.2, 15]$ & $[2,50]$  & \textbf{0.755}  & \textbf{0.751}  & 0.685  & 0.682  & 0.686 & 0.53 \\
       $[1, 5]$ & $[2,100]$  & \textbf{0.759}  &  \textbf{0.734}  & 0.704  & 0.675  & 0.613 & {3.41} \\
        $[1, 15]$& $[2,200]$  & \textbf{0.761}  & \textbf{0.711}  & 0.685  & 0.702  & 0.705 & 7.03 \\
        $[5, 15]$ & $[2,200]$  & \textbf{0.804}  & 0.702  & \textbf{0.716}  & 0.711  & 0.692 & 12.29 \\
       \hline
    \end{tabular}
    \label{table_metrics}
\end{table}
}

{Table~\ref{table_metrics} compares the task reliability of the proposed semi-decentralized MADDPG framework under different metrics and different ranges of $z_{n,m}$ and ${\cal D}_{m}^{t}$, $\forall m, n \in {\cal M}$, where the best and second-best are highlighted, and the relative gains of using LoIU over the second-best metric are provided.
Clearly, LoIU consistently outperforms the other metrics. When the requirements of the update delay and error thresholds are more relaxed (i.e., $E_{n,m}=z_{n,m}\sigma_{n}^2$ and ${\cal D}_{m}^{t}$ are larger), the use of LoIU improves more significantly compared to the other metrics. Under more stringent delay requirements, the performance of AoI can approach that of LoIU, as AoI focuses on the time elapsed. Under stricter error requirements, the use of AoII and AoCI offers closer performance to LoIU, as both of these metrics capture the changes in the robots' {status}es. }

\begin{table}[t]
    \centering
    \caption{ Performance of different metrics under proposed method {($z_{n,m}\in[0.2, 15]$, ${{\cal D}_m^t}\in [2,200]$~milliseconds)}}
    \begin{tabular}{l|c|c|c|c|c}
     \hline
     & LoIU & AoI & AoII & UoI & AoCI \\
    \hline
       Task reliability  & \textbf{0.788}  & 0.713  & 0.704  & 0.693  & 0.705 \\
       Transmission reliability  & \textbf{0.658}  & 0.565  & 0.602  & 0.579  & 0.609 \\
       Avg. time elapsed (ms) & {3.04}  & {3.61}  & \textbf{2.85}  & 3.62  & 3.44\\
       Avg. estimation error  & {5.60}  & 7.18  & \textbf{4.83}  & 6.64  & 6.76 \\
              \hline  
    \end{tabular}
    \label{table_performance}
\end{table}

{Table~\ref{table_performance} shows the task reliability, transmission reliability, the average time elapsed for {status} updates (i.e., $d_{m}^{t}$), and the average estimation error (i.e., $e_{n,m}^{t}$), under the proposed semi-decentralized MADDPG framework and different metrics. 
The average time elapsed is comparable across all metrics. While LoIU does not lead to the lowest average estimation error, it outperforms the other metrics in task reliability and transmission reliability. This advantage arises because LoIU considers the varying thresholds for both time elapsed and estimation errors across different robots. By optimizing the trade-off between these factors, LoIU enhances task and transmission reliability more effectively. In contrast, the other metrics focus primarily on minimizing either latency or estimation error, leading to reduced overall efficiency.}

\section{Conclusion}
In this paper, we have proposed a new LoIU metric for a balanced assessment of the freshness and utility of information needed for robot cooperation, and optimized the schedule of the D2D information transmission in a robot team. We have formulated a Dec-POMDP problem to minimize the average LoIU of a robot team. Using the MADDPG and belief distribution of estimation errors, we have developed a new MADRL-based algorithm to address the problem in a semi-decentralized fashion. Simulations have demonstrated that the proposed LoIU-based semi-decentralized MADDPG improves task reliability by about  {33\%, 44\%, 40\%, and 41\%, compared to AoI-, AoII-, UoI-, and AoCI-based counterparts.} Our approach improves the freshness and utility of cooperation information by up to 98\% compared with these alternatives.
{In future work, we will further explore model compression or lightweight architectures to ensure that the trained models can be efficiently deployed on resource-constrained devices.}
	
    \bibliography{collaborativeMRS}

\end{document}